%% file: Template.tex
\documentclass[10pt, letterpaper]{article}
\usepackage[margin=2cm]{geometry}

\usepackage{amssymb}
\usepackage{amsmath}
\usepackage{graphicx}

\usepackage{cite}

\usepackage{nicefrac}   
\newcommand{\nfrac}[2]{\nicefrac{#1}{#2}}

\usepackage{color}

\newcommand{\dist}[1]{\text{\textbf#1}}
\newcommand{\bulletpoint}[1]{\item[\textbf{#1}]}
\usepackage{url}

\newcommand{\expval}[1]{\langle{#1}\rangle}

\usepackage{rotating}

\title{ 
How many submissions does it take to discover friendly suggested reviewers?
}

\author{Pedro Pessoa$^{1,2}$, Steve Press\'e$^{1,2,3}$ \\
$^1$Center for Biological Physics, Arizona State University,
Tempe, AZ, USA\\
$^2$Department of Physics, Arizona State University,
Tempe, AZ, USA\\
$^3$School of Molecular Sciences, Arizona State University,
Tempe, AZ, USA}
\date{}

\begin{document}

\maketitle
\abstract{
It is ever more common in scientific publishing to ask authors to suggest some reviewers for their own manuscripts. 
The question then arises: How many submissions does it take to discover friendly suggested reviewers?
To answer this question, we present an agent-based simulation of (single-blinded) peer review, followed by a Bayesian classification of suggested reviewers. To set a lower bound on the number of submissions possible, we create a optimistically simple model that should allow us to more readily deduce the degree of friendliness of the reviewer.
Despite this model's optimistic conditions, we find that one would need hundreds of submissions to classify even a small reviewer subset. Thus, it is virtually unfeasible under realistic conditions. This ensures that the peer review system is sufficiently robust to allow authors to suggest their own reviewers.
}

\noindent{\textbf{Keywords}}: 
Peer review,
Simulation, 
Agent-based model,
Bayesian statistics

\newpage

\input{text}

\bibliographystyle{elsarticle-num} 
\bibliography{refs}

\newpage
\input{appendixs}

\end{document}

%% file: text.tex
\section{Introduction}

\paragraph{}
Peer review is the cornerstone of quality control of academic publishing.  However, the daunting task of selecting appropriate reviewers  \cite{Willis16,Fox17} relies in identifying at least two scholars, free of conflict of interest, who have: 1) the necessary expertise to judge the quality and perceived impact; and 2) the willingness to perform the work \emph{pro bono}. On account of this, it is ever more common that journals request, and often require, authors to suggest candidate reviewers. That is, provide names and contact information of scholars the authors deem qualified to review. 

It is natural to imagine, at first glance, that this incentivizes authors to submit ``friendly'' names, implying suggesting reviewers that they have  reason to believe would be favorably inclined toward them. The fear of such peer review manipulation is potentiated by reports that author-suggested reviewers are more likely to recommend acceptance \cite{Schroter06,Wager06,Rivara07,Bornmann10,Moore11,Kowalczuke15,Liang18,Shopovski20}. However, some of these same studies mention that the quality of reports of author-suggested reviewers does not differ from the ones of editor-suggested reviewers \cite{Schroter06,Wager06,Rivara07,Kowalczuke15,Liang18}. It is also reported that the difference in suggesting acceptance by author-suggested and editor-suggested reviewers  is not significant when comparing  reports of the same submission  \cite{Moore11} nor it is observed to have an effect in the article's acceptance \cite{Schroter06,Moore11} and this discrepancy can even vanish entirely in some fields \cite{Zupanc22}. 

The question then naturally arises: can a scientist infer from their personal history of submissions which reviewers are likely to bias the decision in their favor? In what follows, we present an optimistic agent-based model that surely underestimates the number of submissions required to ascertain the friendliness of the reviewer with high confidence. What we find is that, due to multiple sources of uncertainty (\textit{e.g.}, lack of knowledge as to which reviewer the editor selects), such an effort would require a number of submissions vastly exceeding the research output of all but the most productive scientists. That is, hundreds and sometimes thousands of submissions.

As neither a manuscript's submission history, reviewers selected by the editor, nor suggested reviewers by the authors are publicly available, we adapt agent-based simulation models \cite{Bonabeau02,Abar17,Feliciani19}, already used in generating simulated peer review data \cite{Feliciani19}, and develop an inference strategy on this model's output to ask whether we can uncover favorably biased reviewers. This fits into a larger effort to quantitatively study the dynamics of scientific interactions \cite{BARABASI02,Peterson10,Barabasi18,wang21}.

As we initially simulate the data, we intentionally make assumptions using agent-based models that would result in easy classification in order to obtain a lower bound on the number of submissions required to confidently classify reviewers. These assumptions read as follows:
\begin{itemize}
    \bulletpoint{i)} For each submission, the author will always suggest a small number of reviewers (three, in our simulation) from a fixed and small (ten elements, in our simulation) pool of names.
    \bulletpoint{ii)} The editor will always select one of the reviewers suggested by the authors. 
    \bulletpoint{iii)} The ``friendliness'' of any given reviewer remains the same for all subsequent submissions.
    \bulletpoint{iv)} Submissions from the same author all have the same overall quality.
\end{itemize}
Shortly we will lift the assumptions of this ``cynical model" and introduce a ``quality factor model" or simply, quality model. In particular, we will lift assumption iv). As we will see, lifting assumptions will only raise, often precipitously, the already unfeasibly high lower bound on the number of submissions required to confidently classify reviewers and leverage this information to bias reports in their favor.

\section{Methods}\label{sec:methods}
\paragraph{}

\begin{figure*}
\centering
    \includegraphics[width=.9\textwidth]{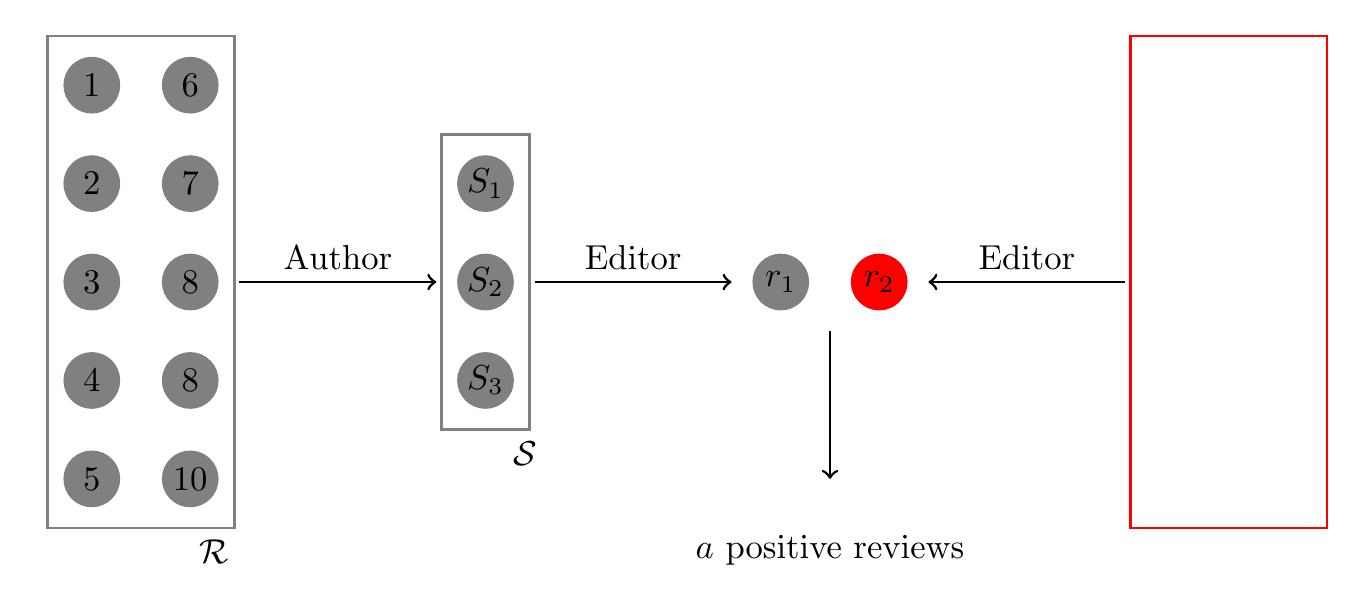}
    \caption{
    Diagram presenting the simplified peer review process. See the steps described in section \ref{sec:methods} for definitions.
    }\label{fig:select_rev}
\end{figure*}

In order to set a lower bound on the number of submissions required to confidently classify reviewers, the present study focuses on a simplified peer review process characterized by three types of agents: the author(s), the editor, and the reviewers. Each submission is reviewed according to the following steps:
\begin{itemize}
    \item[1)] During submission, the author will send to the editor a list of suggested reviewers, $\mathcal{S}$. The suggested reviewers are chosen from a larger set of possible reviewers $\mathcal{R}$ --- such that $\mathcal{S}$ is a subset of $\mathcal{R}$.
    \item[2)] The editor will select one reviewer, namely $r_1$, from $\mathcal{S}$ randomly with uniform probability. 
    \item[3)] The editor will also select a second reviewer, $r_2$, from a pool of reviewers considerably larger than $\mathcal{R}$ and representative of the scientific community.
    \item[4)] The reviewers will write single blind reports, either overall positive or negative, and the author will have access to the number of positive reviews $a$.
\end{itemize}
A diagram of this idealized process is presented in Fig. \ref{fig:select_rev}. 

In the spirit of identifying a lower bound on submissions, we make the dramatic assumption that $r_1$ either belongs to friend or rival class while  $r_2$ is otherwise neutral. Later we will devise a Bayesian inference strategy to achieve suggested reviewer ($r_{1}$) classification.

The procedure described in the bullet points above refers to a single submission. However, as our end goal is to determine how many submissions are necessary to classify reviewers, we must consider multiple submissions. For this reason, we represent a history of $M$, identical and independent, submissions using the index $\mu \in \{1,2,\ldots,M\}$, such that $\mathcal{S}_\mu$ and $a_\mu$ are, respectively, the set of suggested reviewers and positive reviews accrued for the $\mu$-th submission. 

Now that we have qualitatively described our agent-based model, we provide next a detailed mathematical formulation of the simulation and inference.

\subsection{Mathematical formulation}\label{sec:prmodel}
\paragraph{} 

Here each element of $\mathcal{R}$ is a reviewer. We denote $x_i$ the state of each reviewer as belonging to one of two classes: either $x_i=\textit{friend}$ or  $x_i = \textit{rival}$. The method can immediately be generalized to accommodate the addition of a third (neutral) class. Put differently, each suggested reviewer is treated as a Categorical random variable realized to either friend or rival. Collecting all states as a sequence, we write  $x = \left[ x_1, x_2, \ldots, x_{|\mathcal{R}|}\right]$ with $|\mathcal{R}|$ understood as the cardinality of $\mathcal{R}$. For two classes, we have $2^{|\mathcal{R}|}$ allowed configurations of $x$. It is convenient to index configurations with a $j$ superscript where 
$j \in \{ 1,2,\ldots, 2^{|\mathcal{R}|} \} $ for which $x^j = \left[ x^j_1, x^j_2, \ldots, x^j_{|\mathcal{R}|}\right]$.

For sake of clarity alone, we provide a concrete example enumerating all configurations for two possible suggested reviewers in Table \ref{tab:construct_x}.
\begin{table}
\centering
\begin{tabular}{l|lrccll}
j & $x^j$ & $ = [ $ & $x^j_1$ & $x^j_2$ & $]$ &  \\ \hline\hline
1 & $x^1$ & $ = [ $ & \textit{rival},     & \textit{rival}      & $]$ &  \\
2 & $x^2$ & $ = [ $ & \textit{rival},     & \textit{friend}     & $]$ &  \\
3 & $x^3$ & $ = [ $ & \textit{friend},    & \textit{rival}      & $]$ &  \\
4 & $x^4$ & $ = [ $ & \textit{friend},    & \textit{friend}     & $]$ &   
\end{tabular}
\caption{
Example of the construction and enumeration of the possible configurations, $x_j$, for a set of two possible suggested reviewers ($|\mathcal{R}|=2$).  As described in the first paragraph of section \ref{sec:prmodel}.}
\label{tab:construct_x}
\end{table}

We will now use Bayesian inference to determine the probability we assign to each configuration. That is, to compute 
posterior probabilities, $P(x^j|\{a_\mu\},\{\mathcal{S}_\mu\})$, over each $x^j$ given 
the set of positive reports $\{a_\mu\}$ received after suggesting a subset $\{\mathcal{S}_\mu\}$ of reviewers. 
Such inference is only feasible because friend and rival classes exhibit different behaviors when writing reports. In the present article, we will study two models for reviewer behavior.

The first is the, simpler, cynical model where the friend writes a positive review with unit probability and, by contradistinction, the rival writes a positive review with null probability. The reviewer not selected from the author's list, $r_2$, will write a positive review with probability $\nfrac{1}{2}$. In this iteration of the model it should be easiest ({\it i.e.}, quickest in terms of number of submissions) to sharpen our posterior and classify reviewers.

The second model is the quality model that introduces a new layer of stochasticity. Here, a submission is associated a quality factor $q \in (0,1)$ reflecting the quality of each submission . In this model an unbiased reviewer ($r_2$) would write a positive review with probability $q$. By contrast, rivals and friends will  ``double guess'' their own judgment of the article implying that they will evaluate the submitted article twice independently. A rival will only suggest acceptance if they deem the submission worthy of publication in both assessments, meaning a rival will write a positive review with probability $q^2$. Analogously, a friend will reject if they ``reject twice'', hence they write a negative review with probability $(1-q)^2$ or, equivalently, a positive review with probability $1 - (1-q)^2 = q(2-q)$. A summary of these probabilities is presented in Table \ref{tab:quality_accept}. As done with $a_\mu$ and $\mathcal{S}_\mu$, we index the quality factor of the $\mu$-th submission as $q_\mu$.

\begin{table}
    \centering
    \begin{tabular}{c||c|c|c}
         & accept & reject  \\ \hline \hline
        $r_2$ & $q$ &  $1-q$  \\
        $r_1$ is a \textit{rival} & $q^2$  &  $1-q^2$  \\
        $r_1$ is a \textit{friend} & $1-(1-q)^2 = q(2-q)$ &  $(1-q)^2$  \\
    \end{tabular}
    \caption{Probabilities for reviewers of each class to write a positive report (accept) or a negative report (reject) according to the quality model when reviewing a paper of quality factor $q$.  }
    \label{tab:quality_accept}
\end{table}

Not all authors, naturally, have distributions over $q$ centered at the same value. It is therefore of interest to compute the effect on the lower bound of submission needed (\emph{i.e.}, how quickly our posterior sharpens around the ground truth) for different distributions over $q$ centered at the extremes (average high or average low quality) in addition to middle-of-the-road distributions centered at $q=\nfrac{1}{2}$. As we will see, middle-of-the-road distributions allow for more rapid posterior sharpening. Notwithstanding this paltry incentive to write middle-of-the-road papers, we will see that the lower bound on the number of submissions remains unfeasibly high. Even for this idealized scenario.

\subsection{Simulation}\label{sec:simulation}
\paragraph{}

Following the steps described at the beginning of Section \ref{sec:methods}, the first step of the simulation involves editorial selection from the list of suggested reviewers with  ${|\mathcal{R}| \choose |\mathcal{S}_\mu|}$ possible sets of suggested reviewers possible, or $120$ given our simulation parameters ($|\mathcal{S}_\mu| = 3$ and $|\mathcal{R}| =10$  for all $\mu$). Each $\mathcal{S}_\mu $ for any $\mu$ is independently sampled with uniform probability.

We must initialize the ground truth configuration (the identity of $x$). Initially, we set an equal number of friends and rivals though we generalize to two other cases (seven and nine friends) in the Supplemental Information \ref{SIsec:more_friends}.

The subsequent steps (steps 2-3) are straightforward. Step 4 for the cynical model is equally straightforward (and deterministic in $r_1$): 
a positive review is returned if ${r_1}_\mu$ is a friend, a negative review s returned otherwise, while $r_2$ writes a positive review with probability $\nfrac{1}{2}$. Further mathematical simulation details are found in Supplemental Information \ref{SIsec:cynical_samp}.

For the quality model, to each submission ($\mu$) is associated a quality factor $q_{\mu} \in (0,1)$. 
As is usual for a variable bounded by the interval $(0,1)$, we we take $q_{\mu}$ as a Beta random variable such that 
\begin{equation}\label{betadef}
    P(q_{\mu}) =  \frac{q_{\mu}^{\alpha-1}(1-q_{\mu})^{\beta-1}}{B(\alpha,\beta)} 
\end{equation}
where $B(\alpha,\beta)= \frac{\Gamma(\alpha)\Gamma(\beta)}{\Gamma(\alpha+\beta)}$ where $\Gamma$ being the Euler's gamma function. 
Again, in an effort to compute a lower bound alone on the number submissions required, we assume that all $q_\mu$ are sampled from the same, stationary, distribution with constant $\alpha=12$ and $\beta=12$ for now (middle-of-the-road quality distribution) for which the mean $\expval{q} = \nfrac{1}{2}$ and the variance is $\sigma_q=.01$. 

In reality, it is conceivable that one's quality factor distribution shifts to the right with experience. It is also conceivable that a prolific researcher would have its quality factor shift to the left as they start venturing into new fields. This effect only makes it harder to assess which reviewer is friendly and further raises the lower bound required on the number of submissions. In any case, in the Supplemental Information \ref{SIse:diff_betas}, we consider different quality distributions (both high and low). Foreshadowing the conclusions, it may be intuitive to see that very high or very low quality factors result in less information gathered per reviewer report. That is, we learn best the class to which reviewers belong by sampling quality factors around \nfrac{1}{2}. Not by constant rejection or acceptance.

Thus, with each sampled $q_\mu$, step 4) of the quality model is implemented by observing that reviewers write positive reviews according to the probabilities in Table \ref{tab:quality_accept}. Further mathematical details of the quality model are relegated to Supplemental Information \ref{SIsec:quality_samp}. 

Importantly, for the purposes of classifying which reviewers are friendly, it is not necessary to know whether the article is accepted by the editor, only the count of positive or negative reviews per submission.

\subsection{Inference strategy}\label{sec:inference}
\paragraph{}

Inference consists of constructing the posterior $P(x^j|\{a_\mu\},\{\mathcal{S}_\mu\})$ and drawing samples from it. To construct this posterior, we update the likelihood, $P(\{a_\mu\} | x^j,\{ \mathcal{S}_\mu \}$ ), over all independent submission 
\begin{equation}
P(\{a_\mu\} | x^j,\{ \mathcal{S}_\mu \} ) =  \prod_{\mu} P(a_\mu|x^j,\mathcal{S}_\mu)
\end{equation}
as follows
\begin{equation}
    \label{Bayes_theorem}
    P(x^j|\{a_\mu\},\{\mathcal{S}_\mu\}) = \frac{P(x^j|\{\mathcal{S}_\mu\})}{P(\{a_\mu\}|\{\mathcal{S}_\mu\})} \ 
    P(\{a_\mu\} | x^j,\{ \mathcal{S}_\mu \} ) \ .
\end{equation}
Since the number of configurations is finite, we may start by taking the prior as uniform over these countable options ($P(x^j|\{\mathcal{S}_\mu\})=2^{-|\mathcal{R}|}$).  
Keeping all dependency on $x^j$ explicit, we may write
\begin{equation} \label{independence}
    P(x^j|\{a_\mu\},\{\mathcal{S}_\mu\})  \propto P(\{a_\mu\}|x^j,\{\mathcal{S}_\mu\}) = \prod_{\mu} P(a_\mu|x^j,\mathcal{S}_\mu) \ .
\end{equation}

We end with a note on the likelihood which we compute explicitly by treating ${r_1}_\mu$ as a latent variable over which we sum. That is,
\begin{equation}\label{maginalize_r1}
    P(a_\mu|x^j,\mathcal{S}_\mu)  = \sum_{{r_1}_\mu} P(a_\mu|{r_1}_\mu) \ P({r_1}_\mu|x^j,\mathcal{S}_\mu) \ .
\end{equation}
In terms of the factors within the summation, $P({r_1}_\mu|x^j,\mathcal{S}_\mu)$ follows from step 2). That is, if the editor selects ${r_1}_\mu$ with uniform probability from $\mathcal{S}_\mu$, the probability of selecting a ${r_1}_\mu$ from the class of friends is the ratio of friends, $f$, in $\mathcal{S}_\mu$ according to the configuration $x^j$. This can be written more rigorously as
\begin{equation}
     P({r_1}_\mu = \textit{friend} \ |x^j,\mathcal{S}_\mu) = f(x^j,\mathcal{S}_\mu) \doteq \frac{ 1 }{|\mathcal{S}|} \sum_{i\in \mathcal{S}} \  F(x^j_i) ,
\end{equation}
where 
\begin{equation} \label{friend_function}
    F(x^j_i) = 
\begin{cases}
0 & \text{if} \quad x^j_i=\textit{rival} \\
1 & \text{if} \quad x^j_i=\textit{friend} 
\end{cases} \ .
\end{equation}
It follows that  $P({r_1}_\mu = \textit{rival} \ |x^j,\mathcal{S}_\mu) = 1 - f(x^j,\mathcal{S_\mu})$.

We now turn to the term $P(a_\mu|{r_1}_\mu)$ within \eqref{maginalize_r1} computed differently within both the cynical and quality models. 

\subsubsection{Inference in the cynical model}\label{sec:cynical}
\paragraph{}

Calculating $P(a_\mu|{r_1}_\mu)$ for the cynical model is straightforward. That is, given that a friendly $r_1$ always writes a positive review and a rival $r_1$ always writes a negative one, and $r_2$ writes a positive review with probability $\nfrac{1}{2}$, values for $P(a_\mu|{r_1}_\mu)$ immediately follow as tabulated in Table \ref{tab:cynical}. Equations (\ref{Bayes_theorem} -- \ref{friend_function}) and Table \ref{tab:cynical} summarize what is needed to perform Bayesian classification within the cynical model formulation.
\begin{table}[h]
    \centering
    \begin{tabular}{c||c|c|c}
        $P(a_\mu|{r_1}_\mu)$ & $a_\mu=0$ & $a_\mu=1$ & $a_\mu=2$  \\ \hline \hline
        ${r_1}_\mu=$ \textit{friend} & $0$ & $\nfrac{1}{2}$  &  $\nfrac{1}{2}$  \\
        ${r_1}_\mu=$ \textit{rival} & $\nfrac{1}{2}$ & $\nfrac{1}{2}$  &  $0$  \\
    \end{tabular}
    \caption{Probabilities for the number of positive reports, $a_\mu$, in the cynical model, conditioned on the class of the suggested reviewer, ${r_1}_\mu$. 
    }
    \label{tab:cynical}
\end{table}

\subsubsection{Inference in the quality model}\label{sec:quality}
\paragraph{}

The major difference between inference in the quality and cynical models relies on the fact that the author will not have access to individual $q_\mu$'s. However, since we aim for a lower bound, we will proceed with the calculation under the assumption that while individual $q_\mu$'s are unknown the author knows the distribution from which $q_\mu$ is sampled. If the author were uncertain of the distribution, this would add yet another layer of stochasticity and further raise the lower bound.  
From Table \ref{tab:quality_accept}, it is straightforward to calculate the probability of each $a_\mu$ given ${r_1}_\mu$ and $q_\mu$ in the quality model. The result is found in Table \ref{tab:quality}.

\begin{table}
    \centering
    \begin{tabular}{c||c|c|c}
        $P(a_\mu|q_\mu,{r_1}_\mu)$ & $a_\mu=0$ & $a_\mu=1$ & $a_\mu=2$  \\ \hline \hline
        ${r_1}_\mu=$ \textit{friend} & $1-3q_\mu+3q_\mu^2-q_\mu^3$ & $3q_\mu-5q_\mu^2+2q_\mu^3$  &  $2q_\mu^2-q_\mu^3$  \\
        ${r_1}_\mu=$ \textit{rival} & $1-q_\mu-q_\mu^2+q_\mu^3$ & $q_\mu+q_\mu^2-2q_\mu^3$  &  $q_\mu^3$  \\
    \end{tabular}
    \caption{Probability for the number of positive reviews $a_\mu$ conditioned on the quality factor $q_\mu$ and the class of the reviewer ${r_1}_\mu$.
    }
    \label{tab:quality}
\end{table}

Without access to $q_\mu$ in \eqref{maginalize_r1}, we further need to marginalize $P(a_\mu|q_\mu,{r_1}_\mu)$ over $q_\mu$ as follows
\begin{equation} \label{marginalized}
    P(a_\mu|{r_1}_\mu) = \int \mathrm{d} q_\mu \ P(a_\mu,q_\mu|{r_1}_\mu) = \int \mathrm{d} q_\mu \ P(a_\mu|q_\mu,{r_1}_\mu) P(q_\mu) = \bigg\langle  P(a_\mu|q_\mu,{r_1}_\mu)  \bigg\rangle_{q_\mu} \ .
\end{equation}
For example, if $q_\mu$ is sampled from a Beta distribution \eqref{betadef} with parameters $\alpha =\beta=12$, as proposed in Section \ref{sec:simulation}, marginalization \eqref{marginalized} yields the values of $P(a_\mu|{r_1}_\mu)$ shown in Table \ref{tab:quality_marginal}.

\begin{table}
    \centering
    \begin{tabular}{c||c|c|c}
        $P(a_\mu|{r_1}_\mu)$ & $a_\mu=0$ & $a_\mu=1$ & $a_\mu=2$  \\ \hline \hline
        ${r_1}_\mu=$ \textit{ rival} & $.38$ & $.48$  &  $.14$  \\
        ${r_1}_\mu=$ \textit{friend} & $.14$ & $.48$  &  $.38$  \\
    \end{tabular}
    \caption{ 
    Probability for the number of positive reviews $a_\mu$ conditioned on the class of the reviewer ${r_1}_\mu$. Calculated by marginalizing $q_\mu$ in Table \ref{tab:quality}, as described in \eqref{marginalized}, for $\alpha=\beta=12$. }
    \label{tab:quality_marginal}
\end{table}

Inference occurs, otherwise, exactly as in the cynical model, thus summarized by Equations (\ref{Bayes_theorem} -- \ref{marginalized}), and Table \ref{tab:quality}.

\section{Results}\label{sec:results}
\paragraph{}
The previous section was focused on constructing the $2^{|\mathcal{R}|}$-dimensional posterior $P(x^j|\{a_\mu\},\{\mathcal{S}_\mu\})$ otherwise difficult to visualize. Since our goal is to determine the number of submissions required  to correctly classify suggested reviewers, we introduce metrics measuring how well the posterior classifies reviewers. Moreover, these metrics ought to be have an assigned value at each submission, and thus be a function of $m$ for each $m \in \{1, 2, \ldots, M \}$. Thus, for a fixed a data set of $M$ submissions, we calculate each metric using the first $m$ submissions for all $m$. 

Each metric is a stochastic function dependent on the dataset (decisions made by reviewers and quality factors sampled) inherited from the variation of the posterior with the data supplied.  For this reason, we consider multiple metric realizations which allow us to compute their mean, median and $50\%$ and $95\%$ credible (or confidence) intervals. Borrowing language from dynamical systems, we refer to these realizations, up to the $m$th submission, as trajectories. 

\subsection{Metrics}\label{sec:metrics}\paragraph{}
The first metric, akin to a marginal decoder obtained for mixture models \cite{Thompson20,Mathews20}, concerns itself with the probability for the class of one specific reviewer $i$. From the posterior over all configurations, we obtain probabilities over the reviewer $i$'s class through marginalization
\begin{equation}
    \begin{split}
    P(x_i= \textit{friend} \ |\{a_\mu \},\{\mathcal{S}_\mu \}) & = \sum_{j} P(x_i^j = \textit{friend} \ |\{a_\mu \},\{\mathcal{S}_\mu \}) \\
    & = \sum_{j} P\left(x^j |\{a_\mu\},\{\mathcal{S}_\mu\}\right) \ F(x^j_i) \ ,
    \end{split}
\end{equation}
where $F$ was defined in \eqref{friend_function}. Equivalently, $P(x_i= \textit{rival}) = 1 - P(x_i= \textit{friend})$.

Thus, the first metric is defined as the marginal probability of reviewer $i$ being a friend based on the results of $m$  papers where reviewer $i$ was suggested
\begin{equation}\label{rho_def}
    \rho_i(m) \doteq P(x_i = friend \ |\{a_\mu\}_{\mu =1:m},\{\mathcal{S}_\mu\}_{\mu =1:m}) \ ,
\end{equation}
where $\{a_\mu\}_{\mu =1:m}$ and $\{\mathcal{S}_\mu\}_{\mu =1:m}$ represents the subset of the $m$ first elements of $\{a_\mu\}$ and $\{\mathcal{S}_\mu\}$ respectively.

The second metric, a global metric, simply compares the maximum {\it a posteriori} (MAP) estimate for $x$ after $m$ submissions, $\bar{x}(m)$,
\begin{equation}\label{map_def}
    \bar{x}(m) \doteq \arg \max_{x^j} \ P\left(x^j|\{a_\mu\}_{\mu =1:m},\{\mathcal{S}_\mu\}_{\mu =1:m}\right) \ ,
\end{equation}
 and compares, element-wise, how $\bar{x}(m)$ differs from the ground truth.
 
The simulated MAP error, while less informative than considering the full posterior, serves as a estimate on the number of submissions necessary to estimate lower bounds (within tolerable error) to classify as a function of $m$ and the number of friends in the original pool of reviewers. More robustness analysis is performed in Supplemental Information \ref{SIsec:more_friends}.

As a third metric, we look for a more general metric for how ``well-classified'' the reviewers are. Following the work of Shannon \cite{Shannon48}, we notice that entropy defined as 
\begin{equation}\label{entropy_def}
    S(m) \doteq 
    -\sum_{x^j} P\left(x^j|\{a_\mu\}_{\mu =1:m},\{\mathcal{S}_\mu\}_{\mu =1:m}\right)  \ \log_2 P\left(x^j|\{a_\mu\}_{\mu =1:m},\{\mathcal{S}_\mu\}_{\mu =1:m}\right) \ ,
\end{equation}
measures, in rough terms, how many reviewers are left unclassified\footnote{Base $2$ for the logarithm in \eqref{entropy_def} was chosen because we are dealing with binary classification.}. A mock example on how entropy works for the classification of 2 reviewers is presented in Table \ref{tab:entropy_explanation}. For more general insight on the role of entropy see {\it e.g.}, Refs. \cite{Jaynes03,Caticha12,Presse13} and references therein.

\begin{table*}[]\centering
    \begin{tabular}{rccccl||c|c|c|c|c}
    j & $x^j$ & $ = [ $ & $x^j_1$ & $x^j_2$ & $]$      & $P(x^j)$ & $P(x^j)$ & $P(x^j)$ & $P(x^j)$  & $P(x^j)$                  \\ \hline \hline
    1 & $x^1$ & $ = [ $ & \textit{rival},     & \textit{rival}      & $]$  & $\nfrac{1}{4}$ & $\nfrac{1}{8}$  & 0 & 0 & 0          \\ 
    2 & $x^2$ & $ = [ $ & \textit{rival},     & \textit{friend}     & $]$  & $\nfrac{1}{4}$ & $\nfrac{1}{8}$ & 0 & 0  & 0          \\ 
    3 & $x^3$ & $= [ $  & \textit{friend},    & \textit{rival}      & $]$   & $\nfrac{1}{4}$ & $\nfrac{3}{8}$ & $\nfrac{1}{2}$ & $\nfrac{7}{8}$ & 1\\ 
    4 & $x^4$ & $= [ $  & \textit{friend},    & \textit{friend}     & $]$   & $\nfrac{1}{4}$ & $\nfrac{3}{8}$ & $\nfrac{1}{2}$ & $\nfrac{1}{8}$ & 0 \\ \hline\hline
    entropy  &\multicolumn{4}{c}{ $- \sum_j P(x^j) \log_2 P(x^j) $ }& & $2$           & $\approx 1.812$& $1$           & $\approx 0.5436$ & 0 \\
    \end{tabular}
    \caption{
    Example for how entropy is to be interpreted. 
    In this mock example for the classification of two reviewers, similar to Table \ref{tab:construct_x}, all probabilities over configurations (first column) are equally likely and thus the entropy is ascribed its maximal value of $2$. 
    The second column shows a case where the first reviewer is not yet classified, but is considerably more likely to belong to one class, hence entropy takes on some value between $1$ and $2$. 
    The third column contains an example where the first reviewer is fully identified, but the probability does not favor any classification for the second reviewer leading to the entropy value of $1$. 
    The fourth column has an example where the first reviewer is fully identified, but the probability favors one classification (rival) for the second reviewer, leading to the entropy value between $0$ and $1$.
    The last column contains an example where one configuration has probability $1$, hence the reviewers are fully classified, leading to $0$ entropy.      
    }
    \label{tab:entropy_explanation}
\end{table*}

The fourth, and final, metric is the third largest marginal posterior, or the posterior for the third  reviewer most likely to be friendly, 
\begin{equation}\label{T_def}
    T(m) \doteq \max_i^3 \rho_i(m) \ ,
\end{equation}
where $\max\limits_i^n$ is the $n$-th biggest element in the set indexed by $i$ and $\rho_i(m)$ is defined in \eqref{rho_def}.
Unlike the first metric, which classifies each reviewer individually, and  second and third metrics, which classify all reviewers in $\mathcal{R}$, this metric classifies a scenario where authors only 
seek to classify a minimum number of suggested reviewers ($|\mathcal{S}_\mu| = 3$ in our simulations). Therefore, whenever we present results for this fourth metric, we show how many publications are required in order to reach the 95\% confidence level. Despite reaching this metric, it is possible to misclassify the third referee; details provided in Supplemental Information \ref{SIsec:agressive}. In the same Supplemental Information we also explore the possibility that suggesting reviewers based on outcomes from prior optimization on previous submissions does not lead to significant reduction in submissions necessary to classify reviewers.

\subsection{Cynical model results}\label{sec:cynical_results}\paragraph{}

\begin{figure*}
\centering
   \includegraphics[width=.9\textwidth]{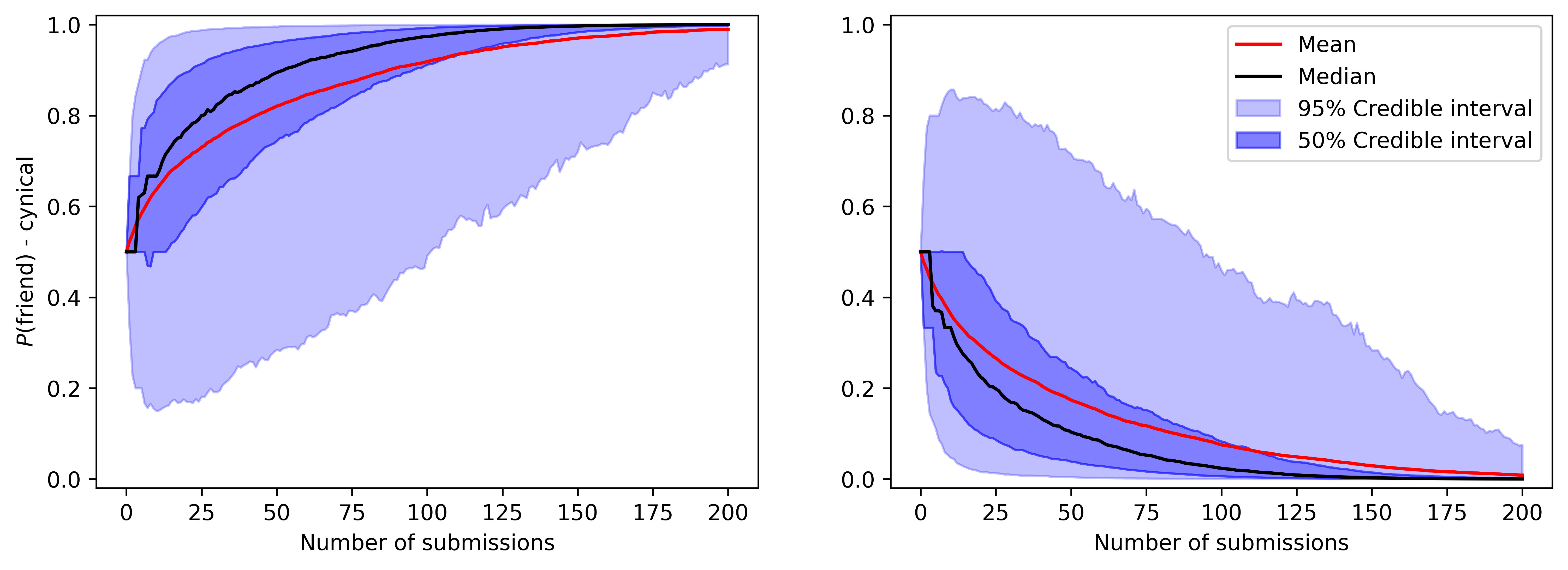}
   \vspace{-.5cm}
    \caption{Posterior marginal probability of a single reviewer's class --- $\rho_i$ defined in \eqref{rho_def} --- as a function of the number of submissions where the reviewer was suggested.
    The graph on the left corresponds to values of $\rho_i(m)$ for which the reviewer belongs to the friendly class in the simulation's ground truth, while the graph on the right corresponds to rivals in the simulation's ground truth. 
    We observe that the median trajectory reaches a probability of $.95$ for the correct class after a little more than $75$ submissions involving the suggested reviewer. 
    Meanwhile, it takes between $100$ to $125$ for the class with the highest posterior to match ground truth within the $95\%$ credible interval for submissions involving this reviewer.
    }
    \label{fig:cynical_prob}
\end{figure*}

The marginal probability (first metric) for the reviewer belonging to the friend class in the cynical model is shown in Fig. \ref{fig:cynical_prob}. We interpret this result as indicating that one needs to suggest this reviewer in a little over than $75$ submissions to strongly classify (marginal posterior exceeding $0.95$ for one of the classes) this reviewer for the median case. Assuming this reviewer is picked uniformly from the author's pool of 10 reviewers then, on average, a total number of 250 submissions would be required.

By contrast, around $100$ submissions suggesting this reviewer are necessary to weakly classify, meaning classify this reviewer using the class that has the highest marginal posterior and obtain the correct class within the  $95\%$ credible interval. In Supplemental Information \ref{SIsec:more_friends} we see that if we have more friends in the ground truth configuration, friends are classified faster, but rivals are likely to be misclassified.

\begin{figure*}    
\centering
  \begin{minipage}[b]{0.45\textwidth}
    \includegraphics[width=\textwidth]{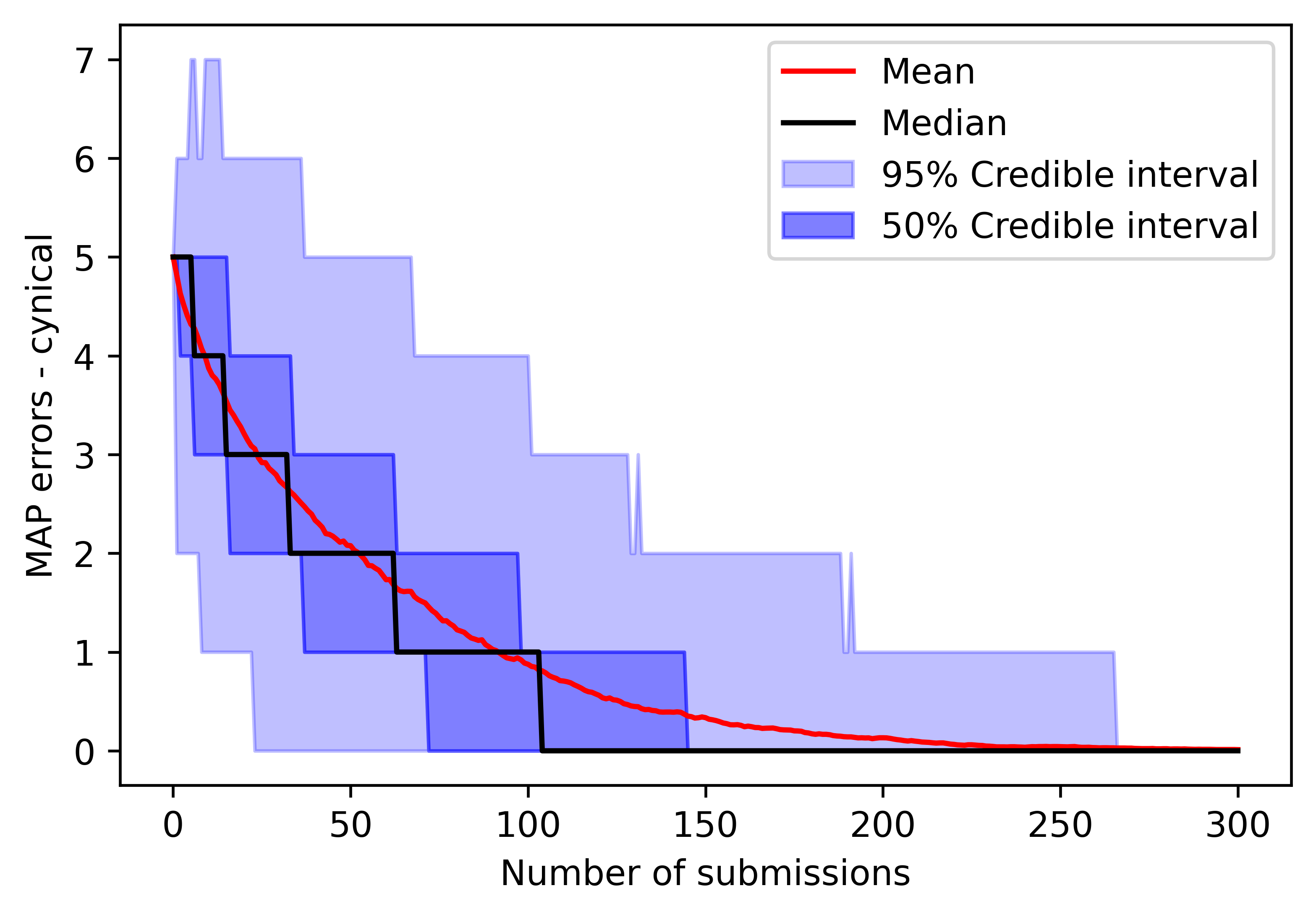}
    \vspace{-.5cm}
    \caption{Number of errors when using maximum {\it a posteriori} (MAP) classification, {\it i.e.}, the number of misclassifications appearing in the MAP configuration \eqref{map_def} when comparing to the simulation's ground truth as a function of the number of submissions in the cynical model.
    We observe that the median trajectory finds the correct ground truth configuration using the MAP estimate after approximately $100$ submissions, while it takes approximately $250$ submissions to reach the correct configuration within the $95\%$ credible interval. }
    \label{fig:cynical_map}
  \end{minipage}
  \hfill
  \begin{minipage}[b]{0.45\textwidth}
    \includegraphics[width=\textwidth]{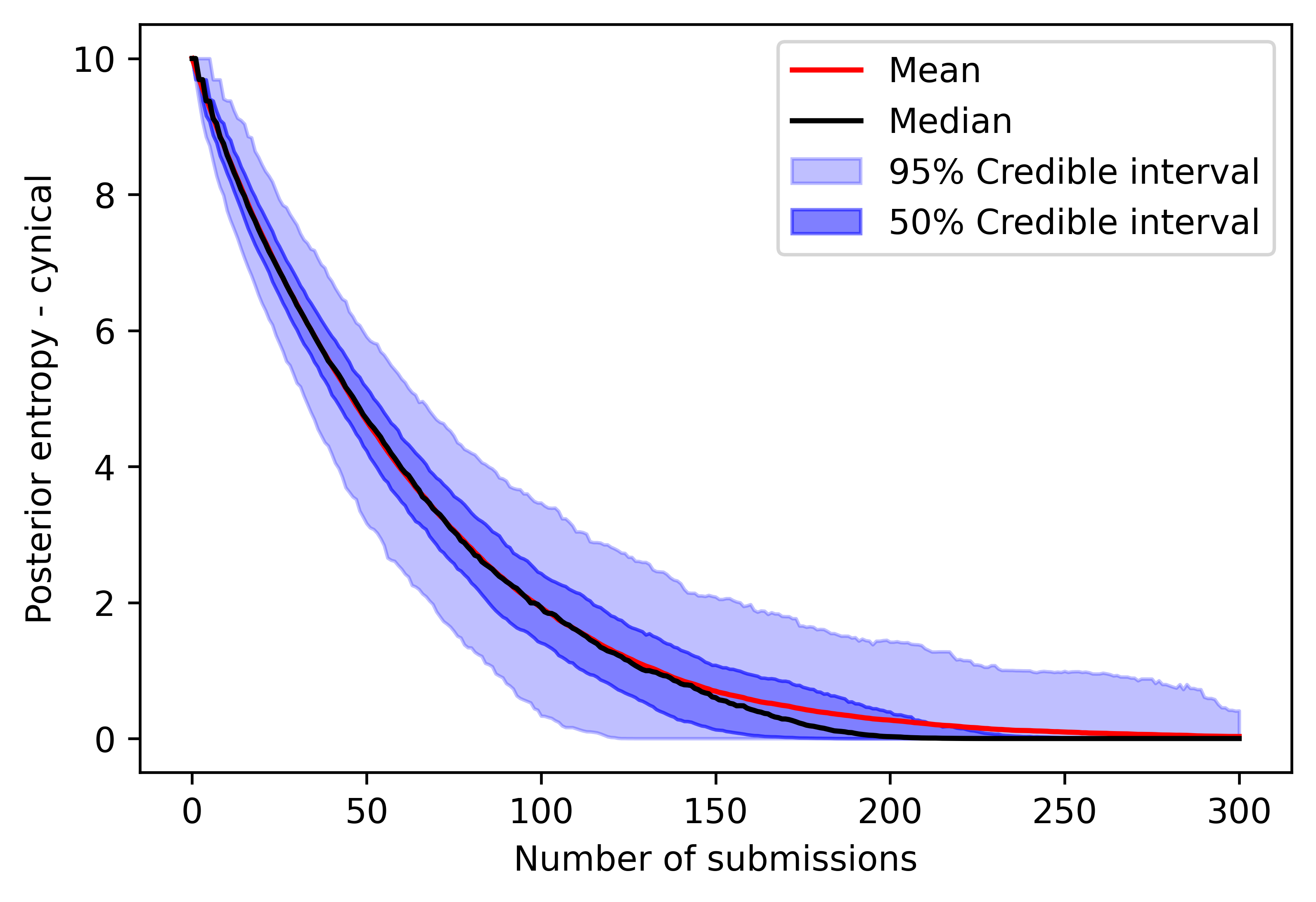}
     \vspace{-.5cm}
    \caption{ 
    The posterior's entropy --- defined in \eqref{entropy_def} --- as a function of the number of submissions in the cynical model. 
    We observe that, in the cynical model, we need between $150$ and $200$ reviewed submissions in order for the entropy of a {median trajectory} to reach zero, meaning that for half of submissions, the posterior only fully classifies a set of 10 reviewers after $150$ submissions.
    }
    \label{fig:cynical_entropy}
  \end{minipage}
   \includegraphics[width=.9\textwidth]{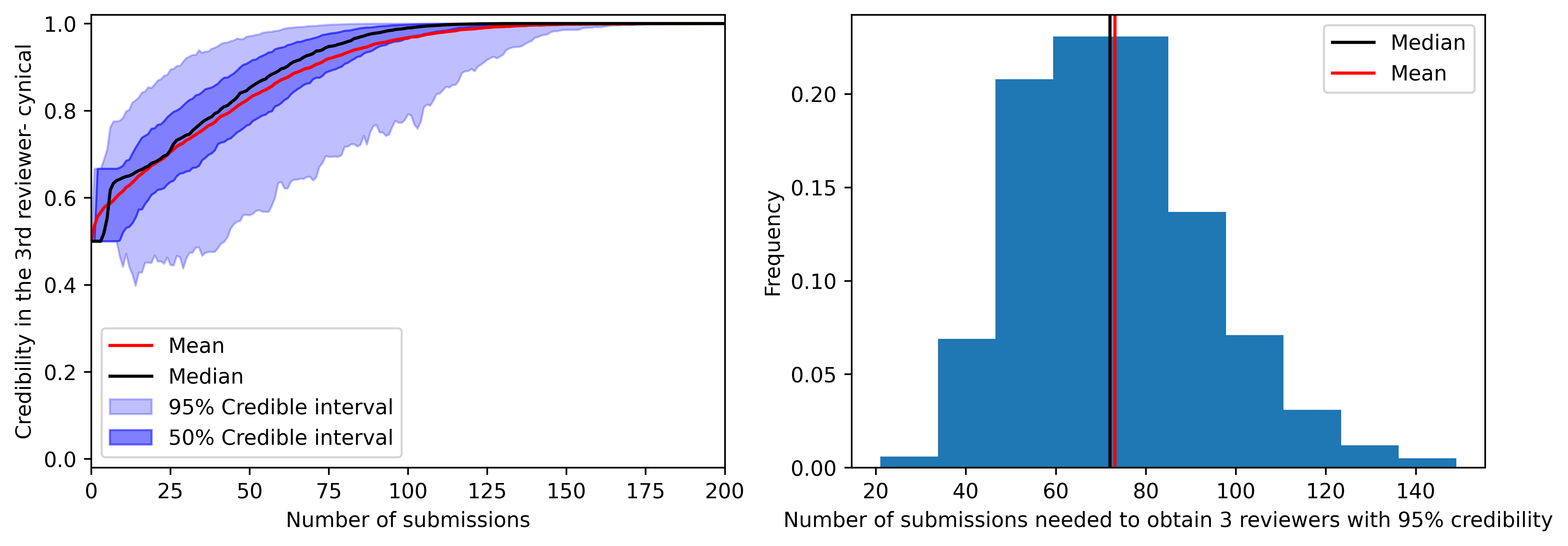}
   \vspace{-.5cm}
    \caption{The left panel presents the marginal probability of the third most likely reviewer (according to the posterior) to be friendly --- defined in \eqref{T_def} ---  as a function of the number of submissions. We observe that the median trajectory's credibility in the third reviewer reaches $95\%$ after approximately $80$ submissions in the median case.
    In the right panel, we see the number of submissions taken to reach $95\%$ credibility for the same metric per sampled simulation. We observe that the mean and median number of submission is slightly bigger than $70$.
    }
    \label{fig:cynical_p3}
\end{figure*}

The number of errors from the MAP (second metric) for the cynical model as a function of the number of submissions is shown in Fig. \ref{fig:cynical_map}. There, we can see that if we attempt to classify reviewers using the MAP, we would get the correct configuration, in the median case, after approximately $100$ submissions. However, to guarantee one finds the correct configuration within the $95\% $ confidence interval, it needs between $250$ to $300$ submissions.

The posterior entropy (third metric) for the cynical model as a function of the number of submissions is shown in Fig. \ref{fig:cynical_entropy}. In this case, we would need, in the median case, between $150$ and $200$ submissions to fully classify a set of $10$ reviewers with $3$ suggested per submission. In the Supplemental Information \ref{SIsec:more_friends}, we see that the posterior entropy does not fall considerably faster (as compared to this case with 5 friends) with more friends in the ground truth.

Finally, we present the third largest marginal posterior as a function of the number of submissions in Fig. \ref{fig:cynical_p3}. We observe that it takes approximately $80$ submissions for the median trajectory to reach $T(m) = 0.95$. In the same figure, we also see that it takes on average $70$ submissions to reach that confidence for all top 3 reviewers. In the Supplemental Information \ref{SIsec:agressive}, we show that if one stops classifying reviewers once they reach that mark, they would classify at least one rival as friend in $6.9\%$ of cases.

\subsection{Quality model results}\label{sec:quality_results}\paragraph{}

\begin{figure*}
\centering
    \includegraphics[width=.9\textwidth]{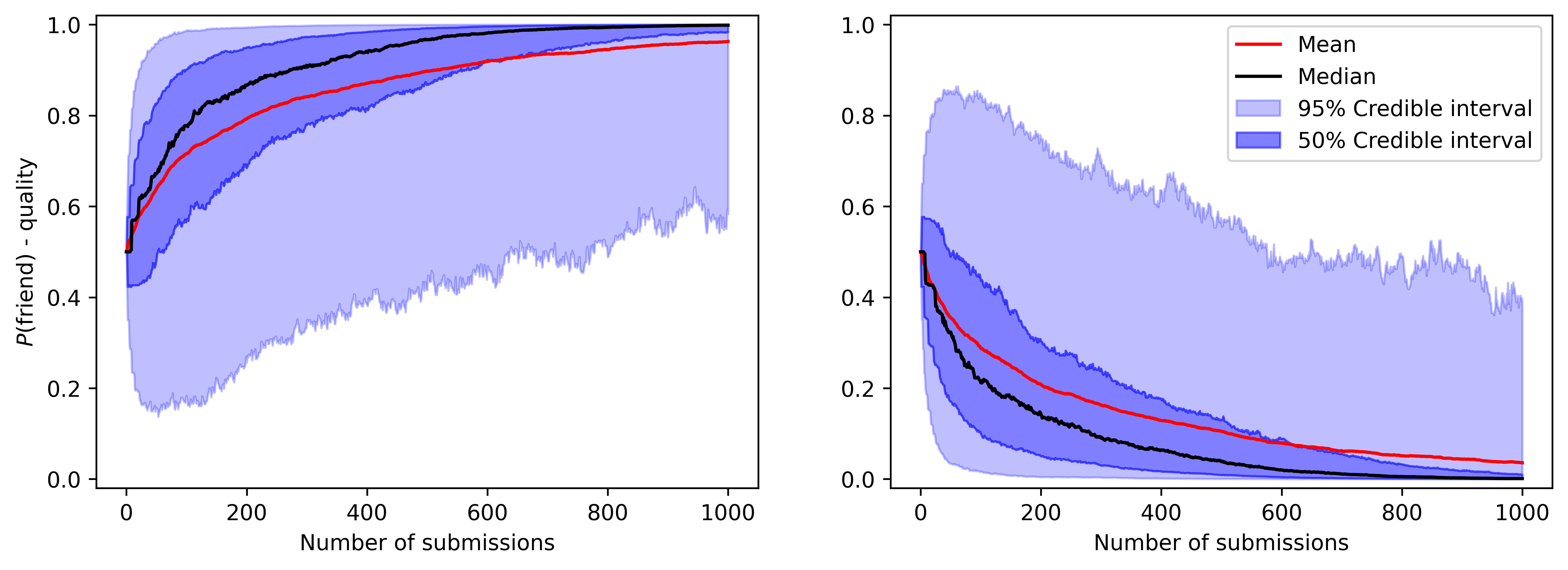}
    \vspace{-.5cm}
    \caption{Marginal posterior probability over a single reviewer's class, analogous to Fig. \ref{fig:cynical_prob},  for the quality model. We observe that {the median trajectory} indicates that  a single reviewer ought to be suggested in approximately $400$ submissions in order to reach a probability of $0.95$ for the correct class. 
    }
    \label{fig:quality_prob}

  \begin{minipage}[b]{0.45\textwidth}
    \includegraphics[width=\textwidth]{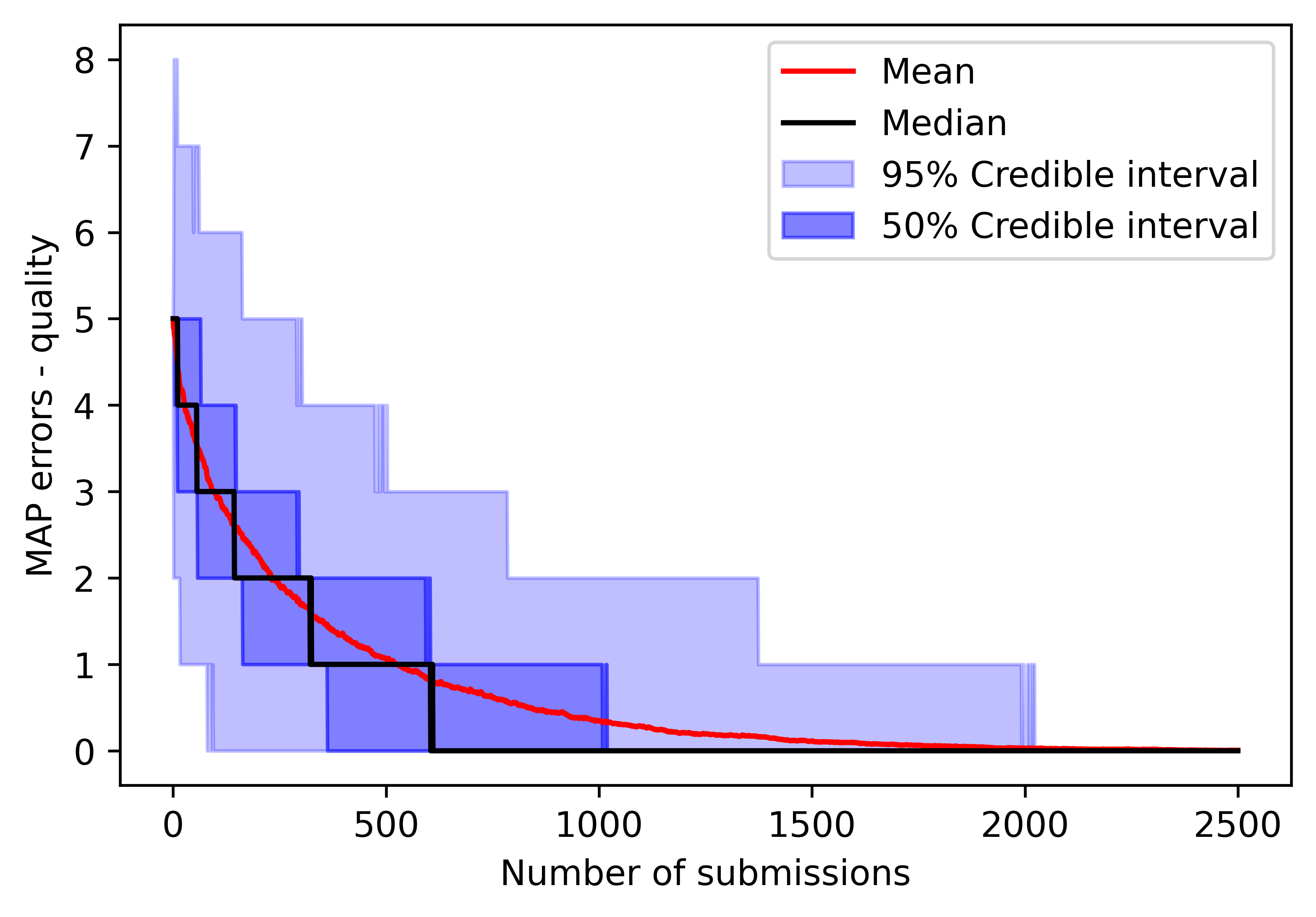}
    \vspace{-.5cm}
    \caption{The number of errors when using maximum {\it a posteriori} (MAP) classification, analogous to Fig. \ref{fig:cynical_map} as a function of the number of submissions. 
    We observe that the median trajectory finds the correct configuration using the MAP estimate after approximately $500$ submissions, while it takes around $2000$ submissions to reach the correct configuration within the $95\%$ credible interval. }
    \label{fig:quality_map}
  \end{minipage}
  \hfill
  \begin{minipage}[b]{0.45\textwidth}
    \includegraphics[width=\textwidth]{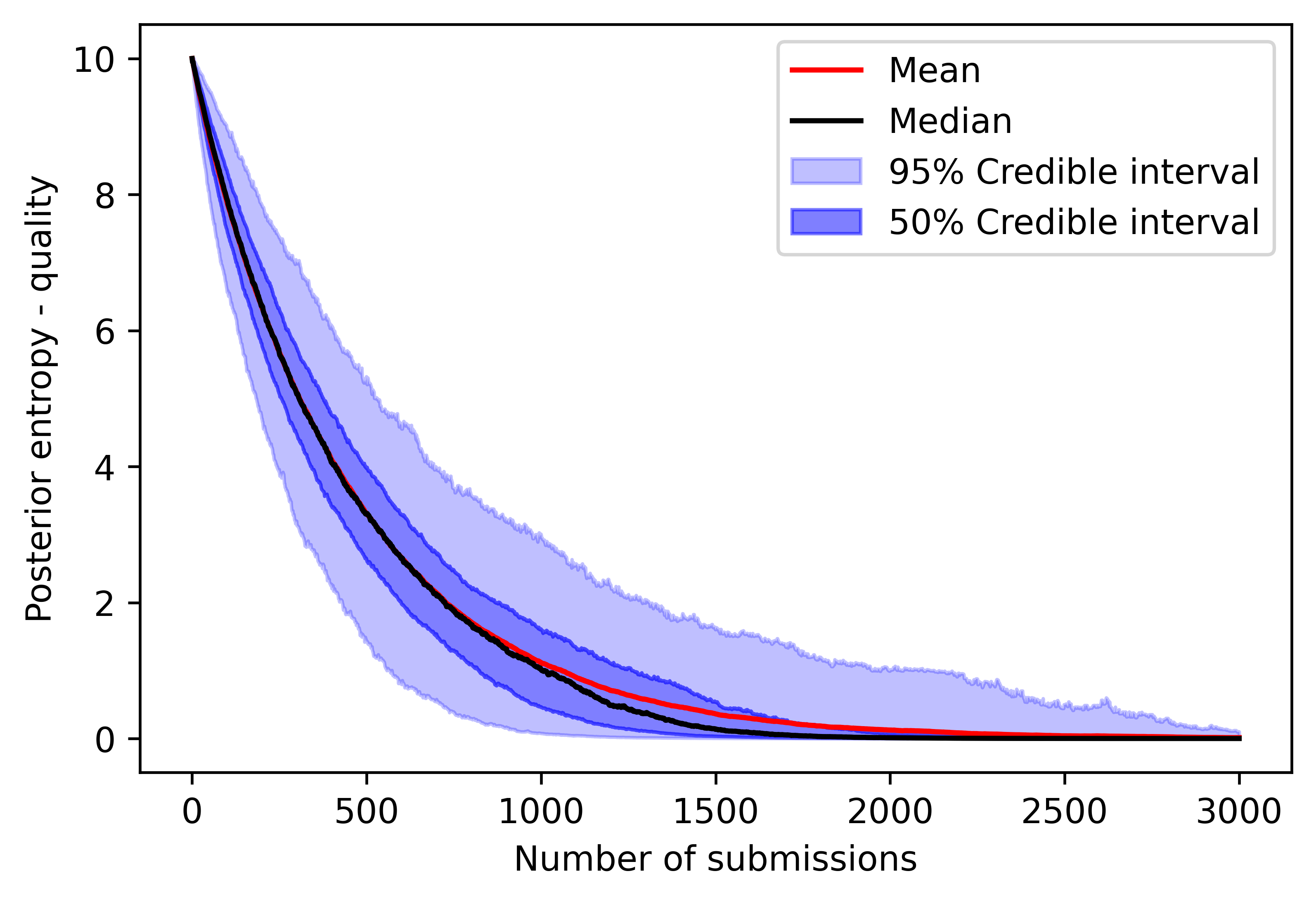}
    \vspace{-.5cm}
    \caption{The posterior's entropy, analogous to Fig. \ref{fig:cynical_entropy}, for the quality model. We observe that we need around $1500$ submissions in order to fully classify the reviewers (entropy approach zero) in the median trajectory.
    }\label{fig:quality_entropy}
  \end{minipage}

     \includegraphics[width=.9\textwidth]{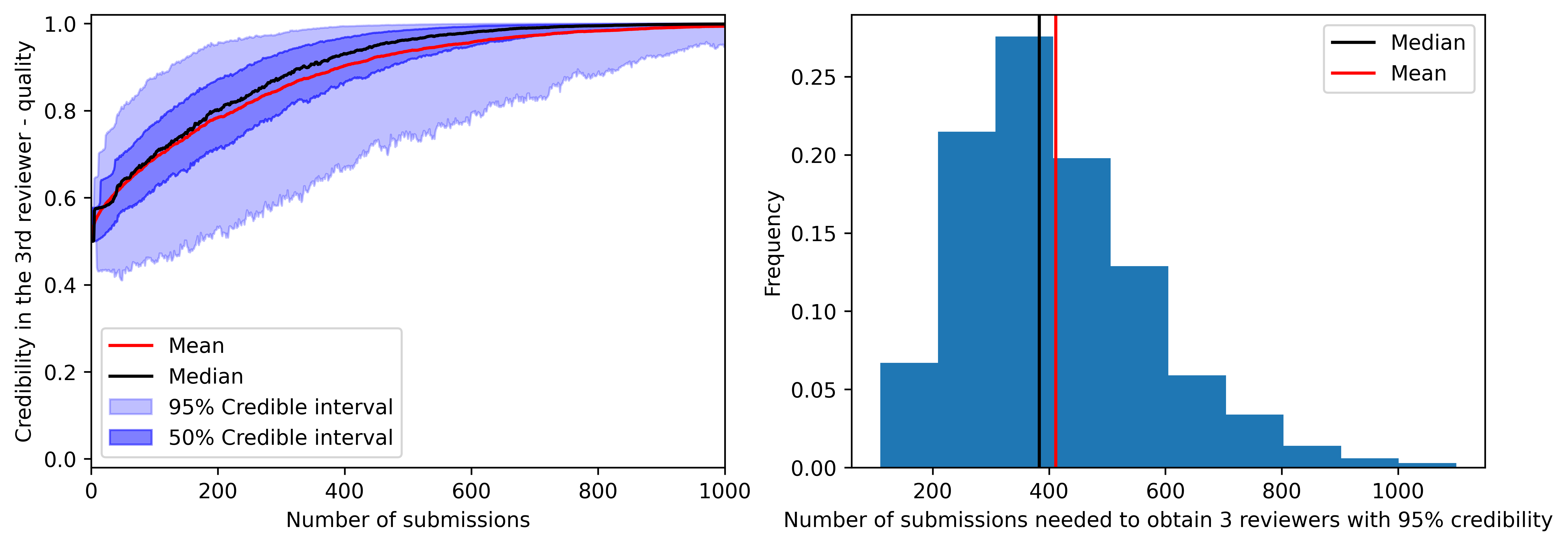}
     \vspace{-.5cm}
    \caption{The left panel presents the marginal probability of the third most likely reviewer to be friendly  as a function of the number of submissions in the quality model, while the right panel presents the number of submissions taken to reach $95\%$ credibility for the same metric, analogous to Fig. \ref{fig:cynical_p3}. Both indicate that approximately 400 submissions are necessary.}
    \label{fig:quality_p3}
\end{figure*}

Similar to the analysis of the cynical model, the marginal probability for a single reviewer class in the quality model is shown in Fig. \ref{fig:quality_prob}. 
The results indicate that one needs to suggest a reviewer on approximately $400$ submissions before they can strongly classify the reviewer in the median case. 

MAP errors as a function of the number of submissions is shown in Fig. \ref{fig:quality_map} indicating that we would need more than $500$ submissions to correctly classify reviewers through MAP in the median case. We would need a little less than $2000$ to find the correct configuration within a $.95$ credible interval.

The posterior entropy as a function of the number of submissions for the quality model is shown in Fig. \ref{fig:quality_entropy}.
The results suggest that we would need more than $1500$ submissions to fully classify a set of $10$ reviewers.

The third largest marginal posterior, as a function of the number of submissions, as well as the number of submissions necessary to reach $95\%$ credibility are presented in Fig. \ref{fig:quality_p3}. We observe that, in the quality model, it takes approximately $400$ submissions to find $3$ friendly suggested reviewers with $95\%$ credibility. On the other hand, in the Supplemental Information \ref{SIsec:agressive}, we show that this misclassifies reviewers in less than $1.0\%$ of datasets.

As mentioned in Sec \ref{sec:simulation}, Fig.  \ref{fig:quality_prob} -- \ref{fig:quality_p3} were constructed in a simulation where the quality factors $q_\mu$ are sampled from a Beta distribution \eqref{betadef} with $\alpha=\beta=12$. We consider other sampling distributions for the quality factor and justify that this unusually tight distribution provides what is close to the overall lower bound in the Supplemental Information \ref{SIse:diff_betas}. For example, any broader distribution ({\it e.g.}, $\alpha=\beta=2$), only further increases the lower bound.

\section{Discussion} \label{sec:discussion}
\paragraph{}

Assessing whether a reviewer is positively or negatively inclined is a question riddled with challenges. For starters, data is not only not publicly available but, as an editorial matter of principle, is kept under lock and key. Yet, the answer to the question posed by the title is not fundamentally unknowable despite the paucity of data. This is because we can simulate, and analyze, realistic outcomes based on agent-based models.

Indeed, doing so, our study shows that it is virtually unfeasible, in a single-blind peer review process, for authors to suggest reviewers that will bias the decision in their favor. 
Even modeling the most cynical and predictable reviewer behavior, we find that an author requires about $100$ submissions to correctly classify even a set of $10$ reviewers, while it takes about $70$ submissions to even find $3$ friendly reviewers with high credibility (see Fig.  \ref{fig:cynical_prob} -- \ref{fig:cynical_p3}). When the model is upgraded to a more realistic one (albeit still too simple), at least 400 submissions become necessary for the same task (see Fig. \ref{fig:quality_prob} -- \ref{fig:quality_entropy}).

This large number exceeds submissions of all but a small minority of even the most prolific scientists. Moreover, large submission numbers introduce further complications. For example, a reviewer may exhibit friendliness toward the author in one area and not another, especially problematic for prolific authors who publish across fields; it is also reasonable to expect that the reviewer may change their opinions in the time necessary to write hundreds of articles.

Further mitigating the severe idealizations of even our marginally more realistic model, would only further compound the difficulty in identifying reviewers. This would be true of any further layer of stochasticity introduced. For example: allow an original pool of reviewers to grow as the author gains more experience in the field; allow the author to suggest a variable number of reviewers (not just 3); allow for neutral suggested reviewers; allow friends to become neutral or rivals over time (or vice versa); allow the author's quality factor distribution to change over time; allow the editor to select a variable number of suggested reviewers.

Naturally, this study assumes that the author tries to identify reviewers using only information available to them. Cases of fraud or collusion should be handled through careful editorial scrutiny. While our simulation assumes an editor that is extremely impartial, a good editor will verify if the suggested reviewers have the necessary competency to properly evaluate the submission,  see {\it e.g.}, the Committee on Publication Ethics (COPE) guidelines \cite{Cope17}. Only after approved by the editor, do reviewers receive invitations. If no candidate is deemed appropriate, editors may very well select no reviewers from the suggested list introducing yet another layer of stochasticity. Therefore the task of finding only the minimal requested number of friendly reviewers is nearly pointless for an author who publishes across fields, as is expected of prolific researchers.

Had a lower bound for the number of reviews found in a cynical model been small, it would have become necessary to consider these complexities in order to identify which, if any, assure the soundness of the single-blind review process. But this is not the case, and the results were even surprising to us. Indeed, even the simplest model confirms that the single-blind review process is sufficiently reliable to allow authors to suggest their own reviewers without clouding or biasing the publication decision.

\section*{Funding}
This work is supported by funds from the National Institutes of Health (grant No. R01GM134426 and R01GM130745).



\section*{Code availability}
\paragraph{} The code performing the simulation, inference, and generating figures is available on GitHub \url{https://github.com/PessoaP/how_many_submissions}

%% file: appendixs.tex
\appendix
\section*{Supplemental Information}

\section{Results with a larger ratio of friendly reviewers}\label{SIsec:more_friends}\paragraph{}

\begin{figure}[p]
    \centering
    \includegraphics[width=.9\textwidth]{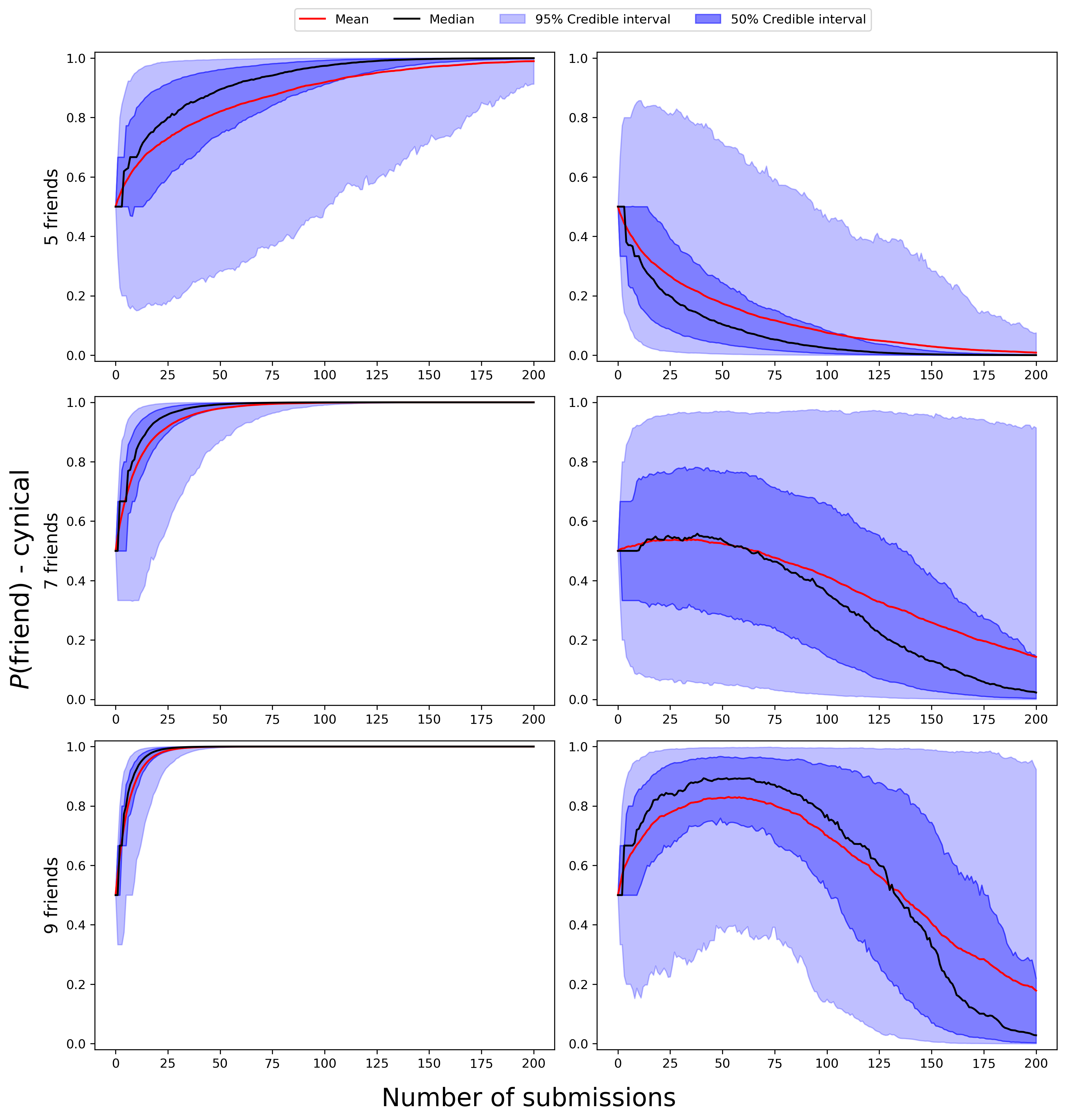}
    \caption{Marginal posterior probability for the cynical model.  
    The figure on the left corresponds to friends in the ground truth configuration and on the left right are rivals in the ground truth configuration. In both cases, these trajectories only take into account submissions where the targeted reviewer was suggested.
    We notice that as the ratio of friends increases, friends are classified with fewer submissions, but rivals are more likely to be misclassified. }
    \label{fig:rho_different_friends_cynical}
\end{figure}

\begin{figure}[p]
    \centering
    \includegraphics[width=.9\textwidth]{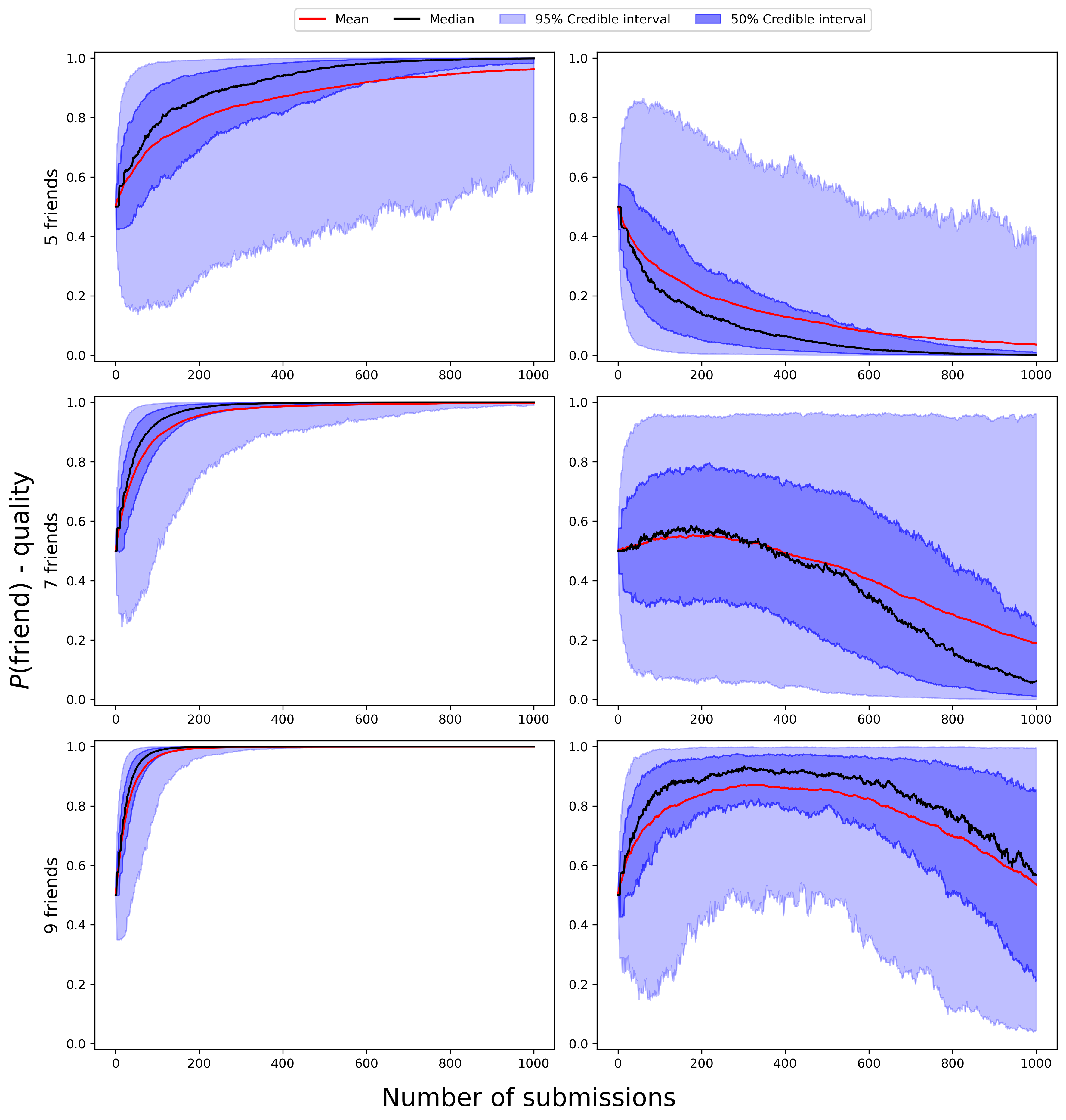}
    \caption{Marginal posterior probability for the quality model.  
     The figure on the left corresponds to friends in the ground truth configuration and on the right are rivals in the ground truth configuration. 
    The pattern is similar to the one in the cynical model, except for the significantly larger number of submissions required. }
    \label{fig:rho_different_friends_quality}
\end{figure}

This supplemental information section presents results for the inference model from simulated data with a different number of friends --- seven and nine out of ten reviewers in $\mathcal{R}$ --- in the ground truth. These are contrasted to the result with five friends presented in the main text. 
Fig. \ref{fig:rho_different_friends_cynical} presents marginal probabilities in the cynical model. We observe that the friendly reviewers are classified faster, but the rivals are likely to be mistaken as friends. This qualitative pattern is repeated in the quality model, as it can be seen in Fig. \ref{fig:rho_different_friends_quality}.

\begin{figure}[p]
    \centering
    \includegraphics[width=.9\textwidth]{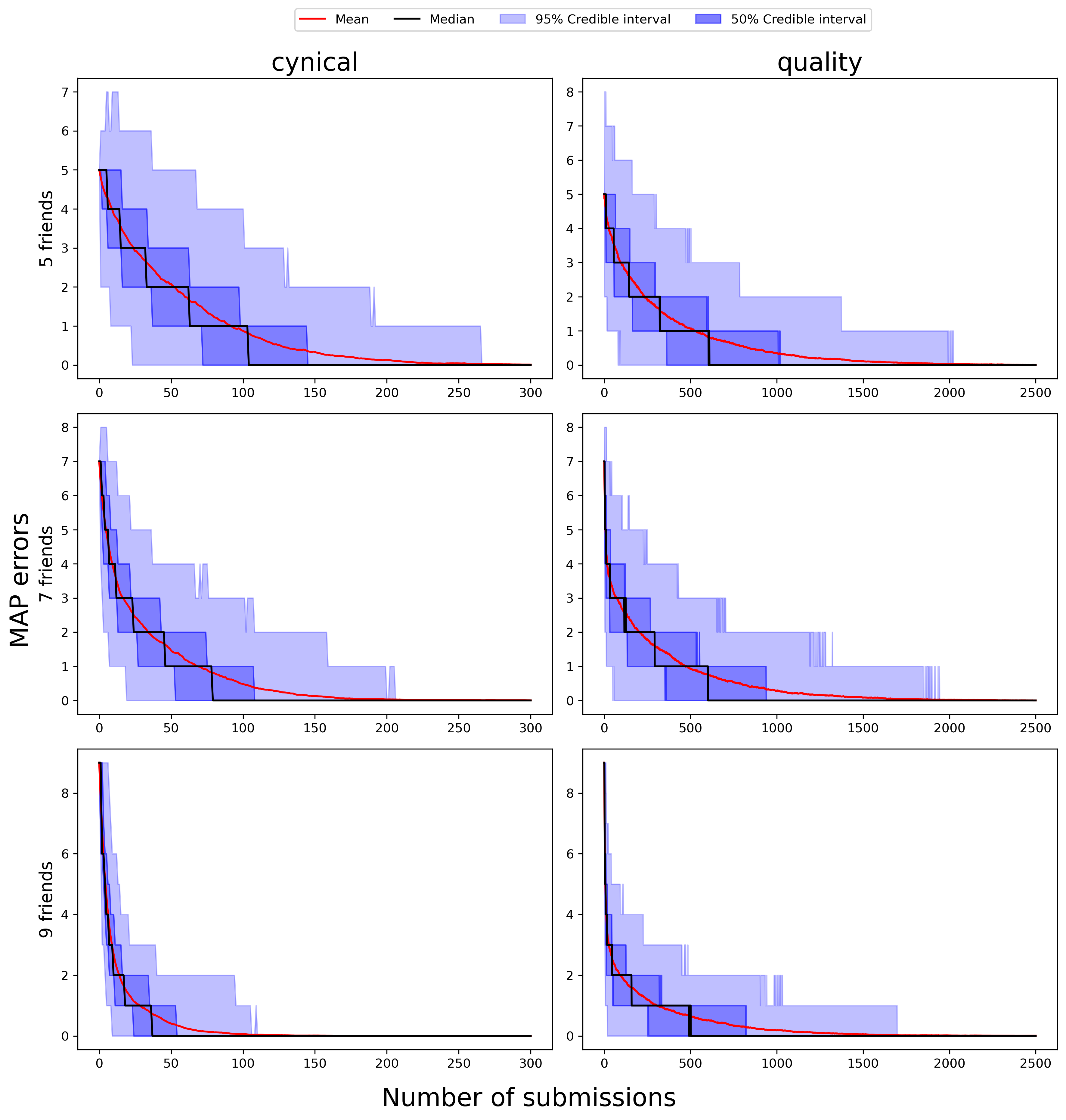}
    \caption{MAP errors for different ratios of friends in the ground truth configuration. 
    We notice that for a greater ratio of friends, it takes fewer submissions for the MAP configuration to match the ground truth configuration. 
    However, this would not necessarily help in classification, as the author does not {\it a priori} know the number of friends in the set of suggested reviewers.}
    \label{fig:map_different_friends}
\end{figure}

Similarly, we see that the MAP classification, presented in Fig. \ref{fig:map_different_friends}, finds the correct configuration with fewer submissions with a larger ratio of friends. However, for a real classification, the ratio of friends is not known {\it a priori}. 
Therefore, it does not necessarily means that an author could classify the reviewers faster.

\begin{figure}[p]
    \centering
    \includegraphics[width=.9\textwidth]{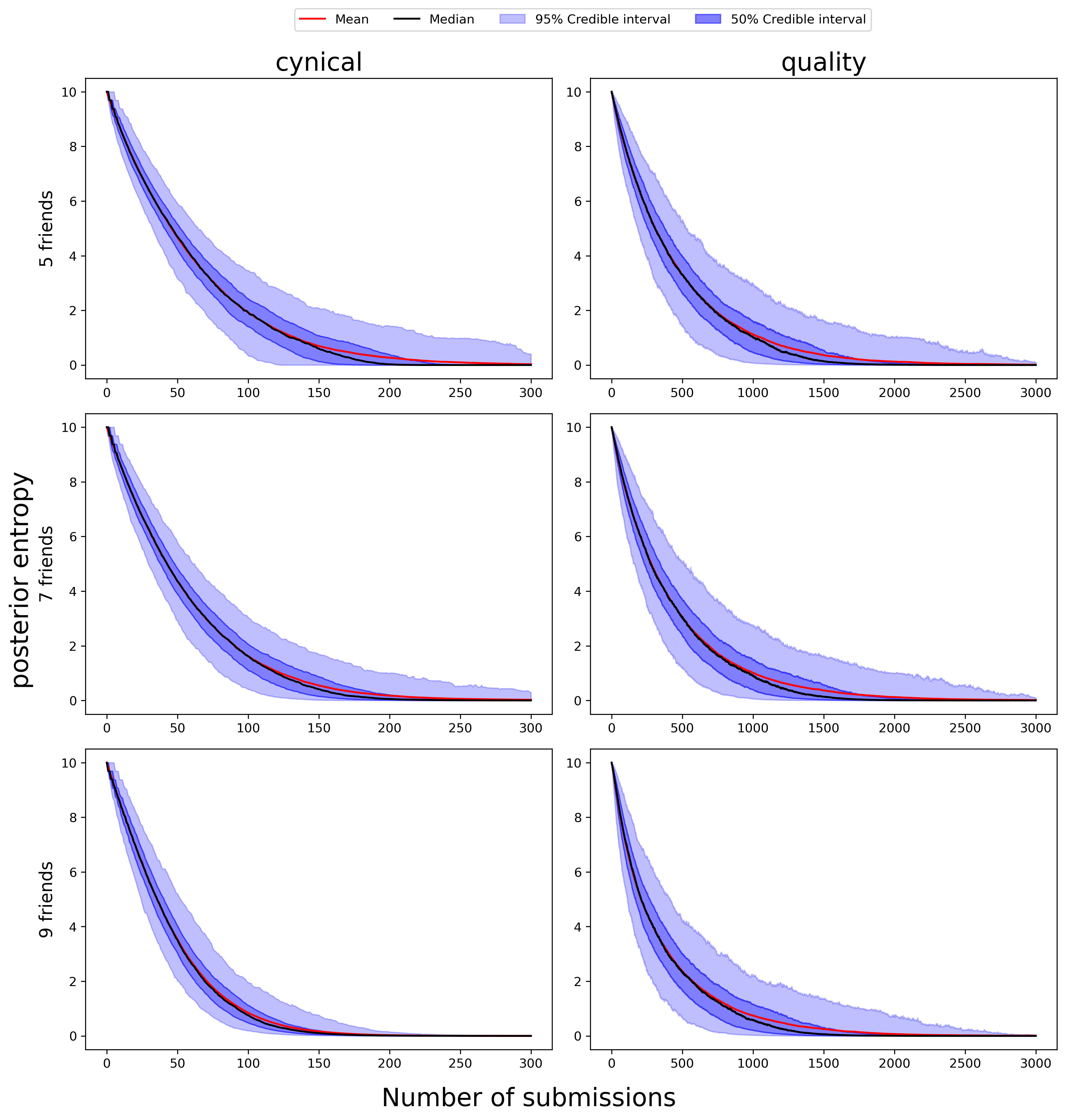}
    \caption{Posterior's entropy for different ratios of friends in the ground truth configuration. The posterior visibly changes with the number of friends in the cynical model. However it is still necessary around 150 submissions to fully classify the reviewers. Nevertheless, the $50\%$ and $95\%$ credible intervals are closer to the median for higher ratio of friends --- indicating smaller fluctuations. In the quality model such changes are not clearly visible. 
    }
    \label{fig:entropy_different_friends}
\end{figure}

Entropy, on the other hand, can be calculated directly from the posterior. So it does not require knowing the number of friends {\it a priori} in order to verify how well classified a set of reviewers is. The posterior entropy for different numbers of friends is presented at Fig. \ref{fig:entropy_different_friends}. We observe that the median number of friends does not change significantly with the ratio of friends, although the fluctuations are smaller in the cynical model. Regardless of the number of friends in the ground truth configuration, it still takes between $150$ and $200$ submissions in the cynical model --- and around $1500$ in the quality model --- to correctly classify a set of ten reviewers. 

\begin{figure}[p]
    \centering
    \includegraphics[width=.9\textwidth]{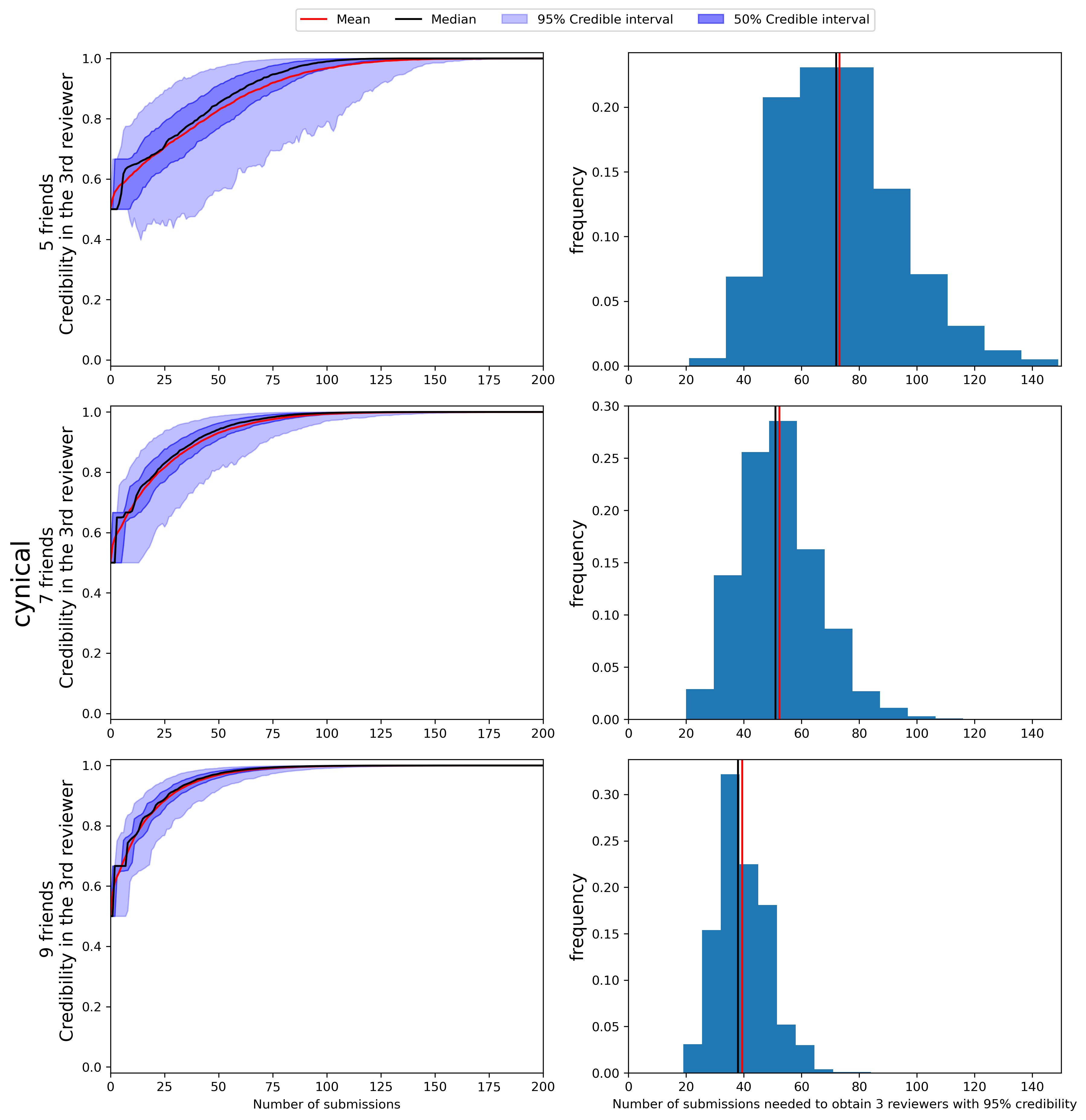}
    \caption{Marginal probability of the third most likely reviewer to be friendly (left) and number of submissions necessary to obtain $95\%$ credibility for three suggested reviewers (right) in the cynical model. The posterior visibly changes with the number of friends. However it is still necessary to have 40 submissions in order to fully classify reviewers. 
    }
    \label{fig:p3_different_friends_cynical}
\end{figure}

\begin{figure}[p]
    \centering
    \includegraphics[width=.9\textwidth]{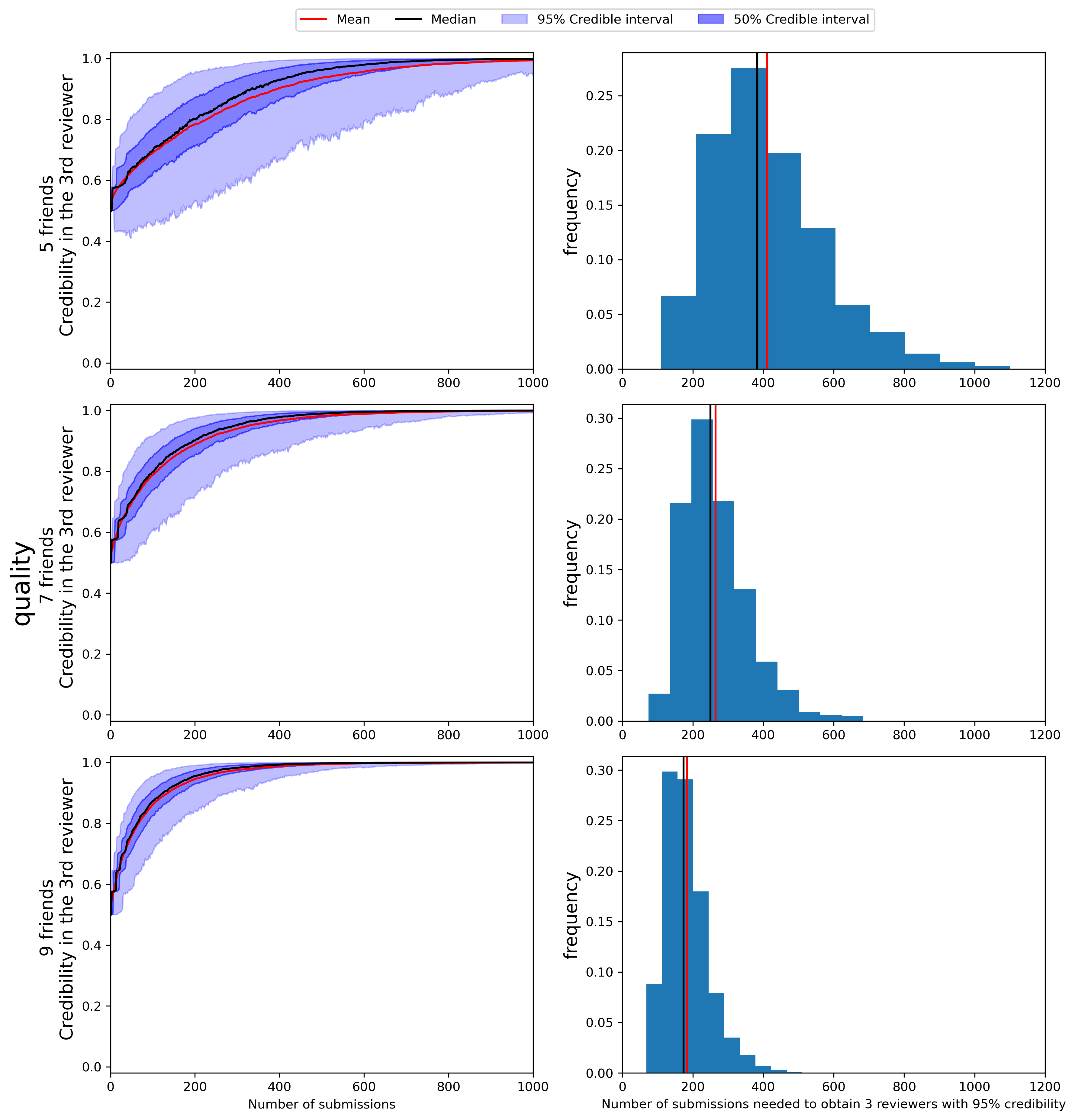}
    \caption{Marginal probability of the third most likely reviewer to be friendly (left) and number of submissions necessary to obtain $95\%$ credibility for three suggested reviewers (right) in the quality model. The posterior visibly changes with the number of friends. However it is still necessary to have around 200 submissions to fully classify the reviewers, a number rather large for all but the most prolific scientists. 
    }
    \label{fig:p3_different_friends_quality}
\end{figure}

Finally, the number of submissions necessary to reach $95\%$ credibility in the cynical model can be seen in Fig. \ref{fig:p3_different_friends_cynical}, as the ratio of friends increase the number of submissions necessary decreases reaching approximately $40$ when there is 9 out of 10 friends. Similarly, as the number of friends in the ground truth increases from $5$ to $9$, it is needed a little less than $200$ submissions to obtain the same credibility in the quality model (Fig. \ref{fig:p3_different_friends_quality}).

\newpage
\section{Sampling}
\paragraph{}
In this supplemental information section, we detail the simulation described in Section \ref{sec:simulation}. 
We present sampling equations for a single submission review, this process is repeated $M$ times and each sampled number of positive reviews, $a$, is assigned a label $\mu \in \{1,2,\ldots,M\}$. 
We separate the sampling equations for the cynical and quality models in the two following subsections.

\subsection{Cynical}\label{SIsec:cynical_samp} \paragraph{}
The simulation generating data in the cynical model begins with the editor choosing one reviewer, $r_1$, from the list of suggested reviewers, $\mathcal{S}$, with uniform probability. 
As done in section \ref{sec:inference}, we use $r_1$ as the class of the suggested reviewer selected by the editor. If reviewer $i$ is selected, $r_1 = x_i$ where $x$ is the ground truth configuration. 
The number of positive reports, $a$ is the sum of two terms: the first, $\chi_{r_1}$ corresponding to the report written by $r_1$, hence it will be $1$ if $r_1 = \textit{friend}$ and $0$ if $r_1 = \textit{rival}$; while the second one, $\chi_{r_2}$ corresponding to the report written by $r_2$, which is equally likely to be $0$ or $1$. The summary of the relevant sampling equations is 
\begin{subequations} \label{Summary_Cynical}
\begin{align}
r_1 &\sim \dist{Categorical}_{s_{1:|\mathcal{S}|}} \left(\frac{1}{|\mathcal{S}|}, \ldots, \frac{1}{|\mathcal{S}|}\right) \ , \\ \label{sampler_cynical}
\chi_{r_1} &= 
\begin{cases}
0 & \text{if} \quad r_1=\textit{rival} \\
1 & \text{if} \quad r_1=\textit{friend} 
\end{cases} \ , \\ 
\chi_{r_2} &\sim \dist{Categorical}_{0,1} \left(\nfrac{1}{2}, \nfrac{1}{2}\right) \ , \\
a &= \chi_{r_1} +\chi_{r_2} \ ;
\end{align}
\end{subequations}
where the symbol $\sim$ means ``sampled from''.

\subsection{Quality}\label{SIsec:quality_samp} \paragraph{}
As mentioned in Section \ref{sec:quality}, the major difference between the quality and the cynical model is that we need to sample the quality factor $q$, using a Beta distribution as in \eqref{betadef}.
As in the cynical model, the number of positive reports is the sum of two terms $\sigma_{r_1}$ and $\sigma_{r_2}$ representing the reviewers $r_1$ and $r_2$ respectively.
In accordance with Table \ref{tab:quality_accept}, if $r_1 = \textit{rival}$, then $\chi_{r_1}$ will be $1$ --- or the report written by $r_1$ will be positive --- with probability $q^2$ and $\chi_{r_1} =0$ with probability $1-q^2$. If $r_1 = \textit{friend}$, we have $\chi_{r_1}=1$ with probability  $q(2-q)$,  $\chi_{r_1}=0$ and with probability $(1-q)^2$. 
Similarly, $\chi_{r_2}$ will be $1$ with probability $q$ and $0$ otherwise.
Also, as in the cynical model, the editor selects the reviewer $r_1$ uniformly from the list of suggested reviewers $\mathcal{S}$. The summary of the relevant sampling equations is 
\begin{subequations} \label{Summary_Quality}
\begin{align}
\label{q_sampler} q & \sim \dist{Beta}(\alpha,\beta)\\
r_1 &\sim \dist{Categorical}_{s_{1:|\mathcal{S}|}} \left(\frac{1}{|\mathcal{S}|}, \ldots, \frac{1}{|\mathcal{S}|}\right) \ , \\ \label{sampler_quality}
\chi_{r_1} &\sim 
\begin{cases}
\dist{Categorical}_{0,1}  (1-q^2,q^2)  &\text{if} \quad r_1=\textit{rival} \\
\dist{Categorical}_{0,1} ((1-q)^2, q(2-q)) &\text{if} \quad r_1=\textit{friend} 
\end{cases} \ , \\ 
\chi_{r_2} &\sim \dist{Categorical}_{0,1} \left(1-q,q\right) \ , \\
a &= \chi_{r_1} +\chi_{r_2} \ .
\end{align}
\end{subequations}

\newpage

\section{Quality results with different parameters}\label{SIse:diff_betas}\paragraph{}
In this supplemental information section, we justify the use of the Beta distribution with parameters $\alpha=12$ and $\beta=12$ to present a lower bound on the number of submissions.  
As mentioned in main text, beta distributions are typical choices for sampling random variables distributed over the interval $(0,1)$. From \eqref{betadef} we calculate the expected value and variance of the quality factor, $\expval{q}$ and $\sigma_q$ respectively, obtaining
\begin{equation}\label{direct}
    \expval{q} = \frac{\alpha}{\alpha+\beta} 
    \quad , \quad
    \sigma_q = \frac{\alpha\beta}{(\alpha+\beta)^2(\alpha+\beta+1)} \ .
\end{equation}
Which can be inverted as
\begin{equation}\label{inverted}
    \alpha = - \expval{q}  \left( \frac{\expval{q}^2 - \expval{q} + \sigma_q}{\sigma_q} \right)
    \quad , \quad
    \beta = \alpha \left( \frac{1}{\expval{q}}-1 \right) \ .
\end{equation}    

In the main text, we studied a scientist with median papers with small variance --- $\alpha =12$ and $\beta=12$ implies, from \eqref{direct}, $\expval{q} = \nfrac{1}{2}$ and $\sigma_q = .01$. Here we will compare this to scientists with  overall smaller and bigger qualities --- $\expval{q} = .25$ and $\expval{q} = .75$ respectively --- and smaller and bigger variances --- $\sigma_q = .05$ and $\sigma_q = .005$ respectively. In Table \ref{SIC-table} we obtain the associated values of $\alpha$ and $\beta$ for this study.
\begin{table}[h]
    \centering
    \begin{tabular}{l||ccc}
            $\expval{q} \backslash \sigma_q$ & .05 & .01 & .005  \\
            \hline \hline
                .25 & $\alpha  = 0.6875$  & $\alpha  = 4.4375$ & $\alpha  = 9.125$  \\
                & $\beta = 2.0625$ & $\beta = 13.3125$   &  $\beta = 27.375$ \\ \hline
            .5 & $\alpha = 2$ & $\alpha = 12$ & $\alpha = 24.5$  \\
               & $\beta = 2 $ & $\beta = 12 $ & $\beta = 24.5 $  \\ \hline
            .75 & $\alpha = 2.0625$ & $\alpha = 13.3125$ & $\alpha = 27.375$  \\
                & $\beta  = 0.6875$ & $\beta  = 4.4375$  & $\beta  = 9.125$  \\ \hline
        \end{tabular}
        \caption{Values of $\alpha$ and $\beta$ obtained from \eqref{inverted} in terms of $\expval{q}$ and $\sigma_q$ for the examples we study in this section. }\label{SIC-table}
\end{table}

\begin{sidewaysfigure}
    \centering
    \includegraphics[width=.8\textwidth]{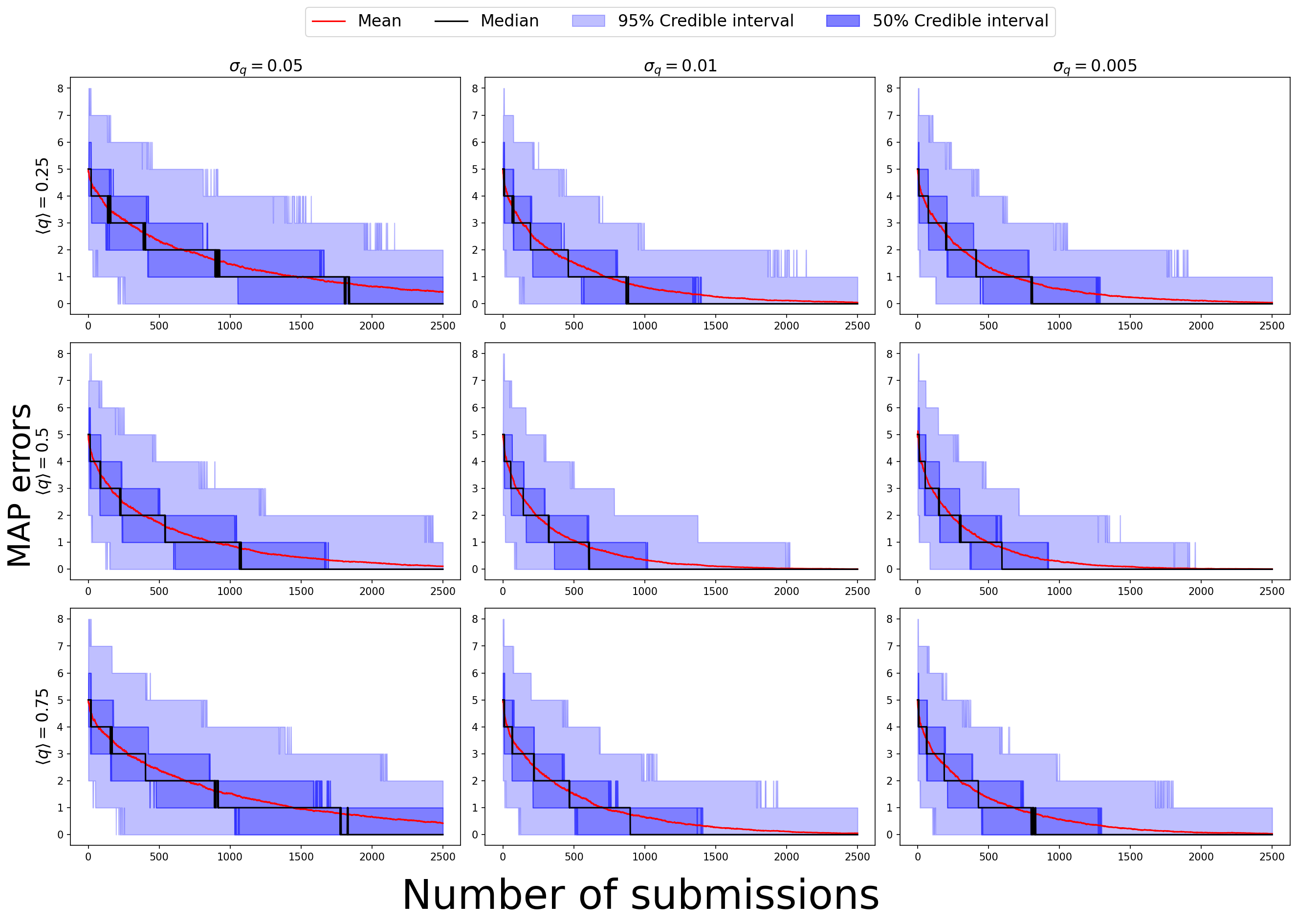}
    \caption{MAP errors obtained for different sampling distributions of quality factors. 
    We notice that the number of submissions necessary to find the correct configuration is smaller for the  expected value $\expval{q} = .5$  --- representing a scientist with overall median quality articles. 
    The number of submissions necessary also decreases as the variance reduces.
    }
    \label{fig:C_map_different}
\end{sidewaysfigure}

\begin{sidewaysfigure}
    \centering
    \includegraphics[width=.8\textwidth]{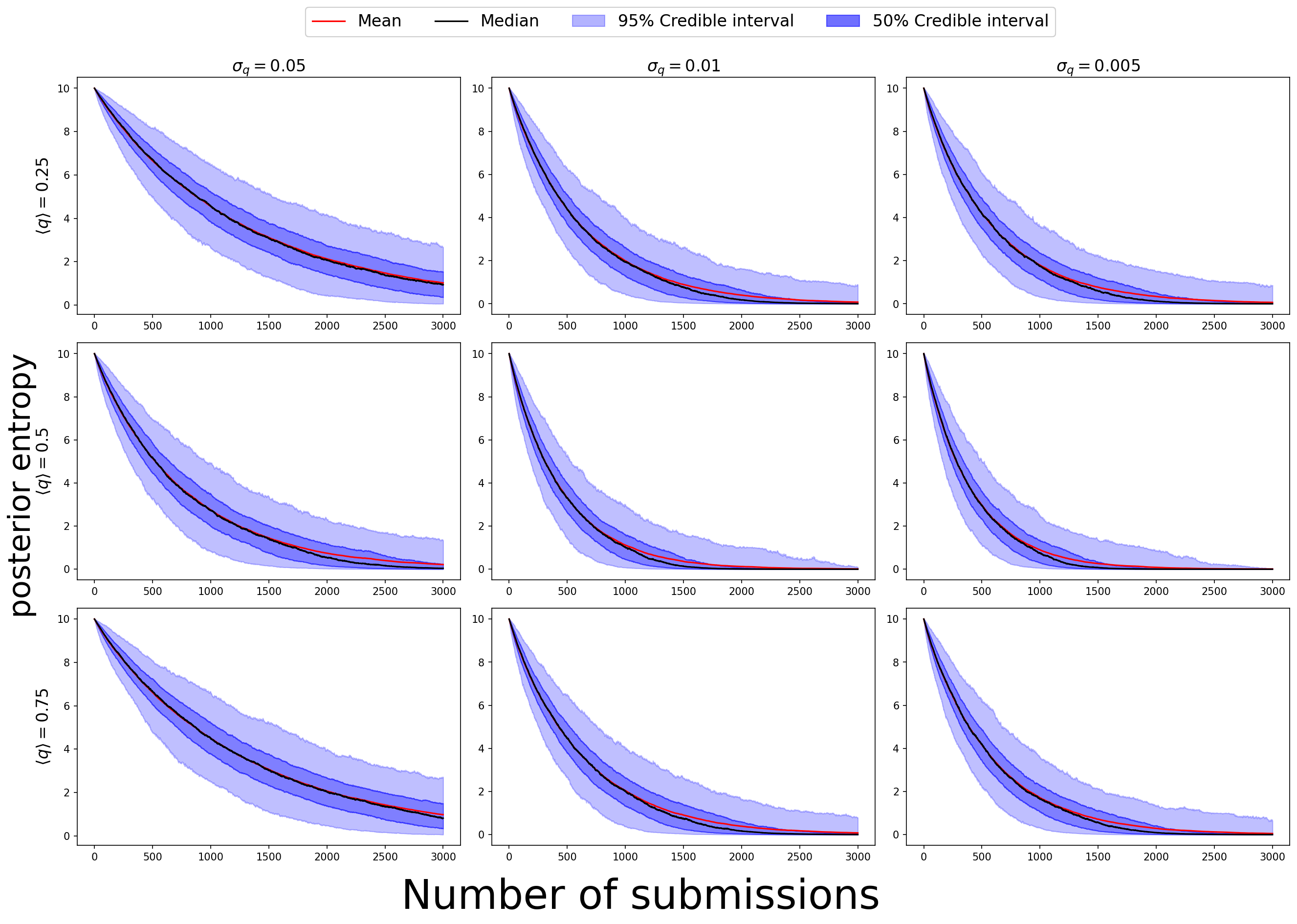}
    \caption{The posterior's entropy obtained for different sampling distributions of quality factors. 
    We notice that the number of submissions necessary to fully classify reviewers is smaller for the expected value $\expval{q} = \nfrac{1}{2}$ when compared to both larger and smaller expected values. 
    The number of submissions necessary also decreases as the variance reduces.
    }
    \label{fig:C_entropy_different}
\end{sidewaysfigure}

Fig. \ref{fig:C_map_different} presents the MAP errors for each of these values. 
We observe that, as the values of $\expval{q}$ move away from $\nfrac{1}{2}$, it require more submissions to correctly classify reviewers. Thus, the lower bound is found by a researcher whose submissions are of median quality $\expval{q} = \nfrac{1}{2}$.  These results are confirmed by an equivalent figure plotting the posterior's entropy in Fig.
\ref{fig:C_entropy_different}.

\begin{sidewaysfigure}
    \centering
    \includegraphics[width=.8\textwidth]{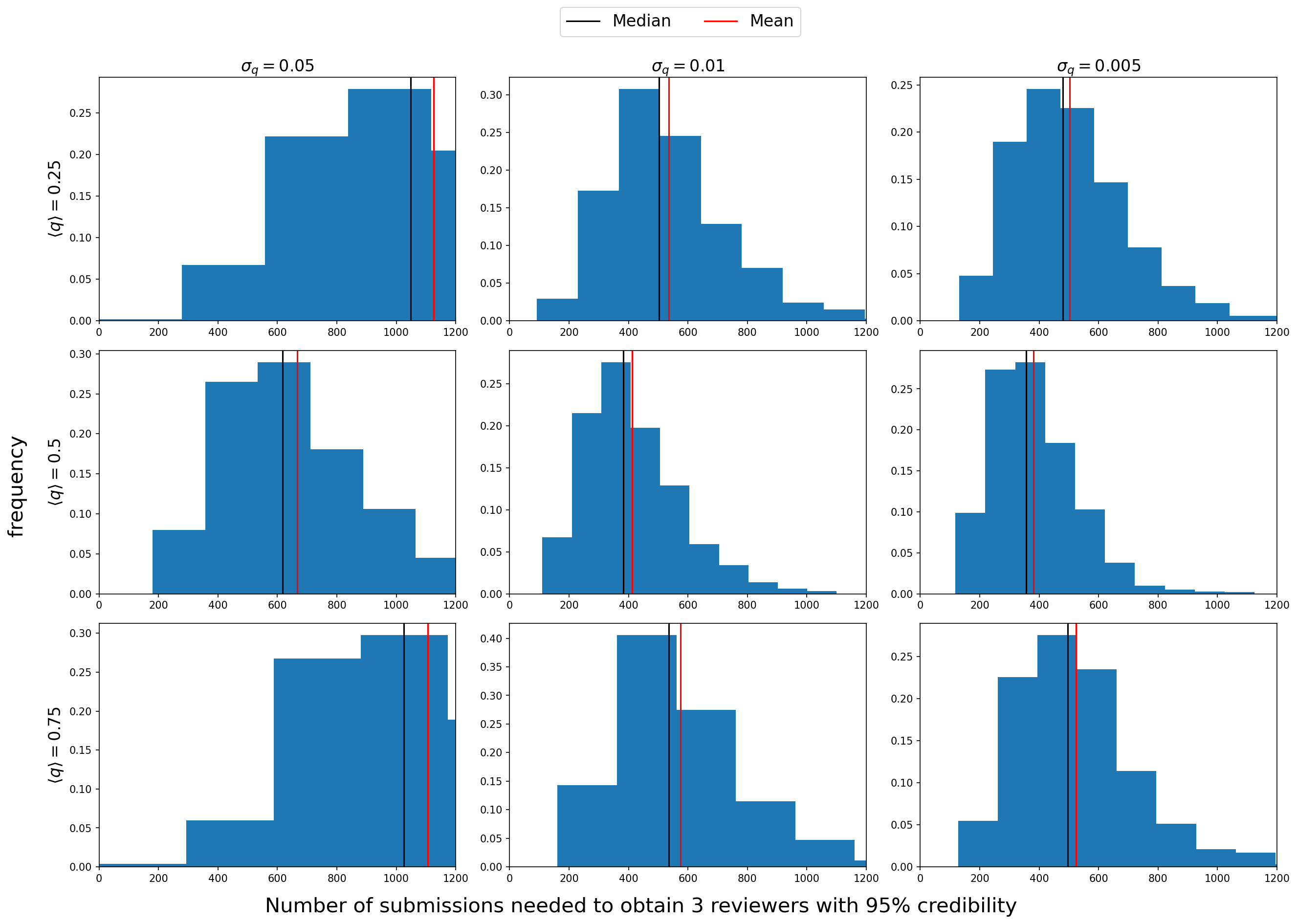}
    \caption{Number of submissions necessary to achieve $95\%$ credibility for three suggested reviewers obtained for different  quality factor distributions. 
    Some of these graphs were truncated in order to present better comparison to the central case, $\expval{q} = \nfrac{1}{2}$.
    We notice that the number of submissions necessary to classify at least three suggested reviewers is smaller for the expected value $\expval{q} = \nfrac{1}{2}$ when compared to both larger and smaller expected values. The number of submissions necessary also decreases as the variance reduces. 
    }
    \label{fig:p3_different}
\end{sidewaysfigure}

\begin{figure}[p]
    \centering
    \includegraphics[width=.9\textwidth]{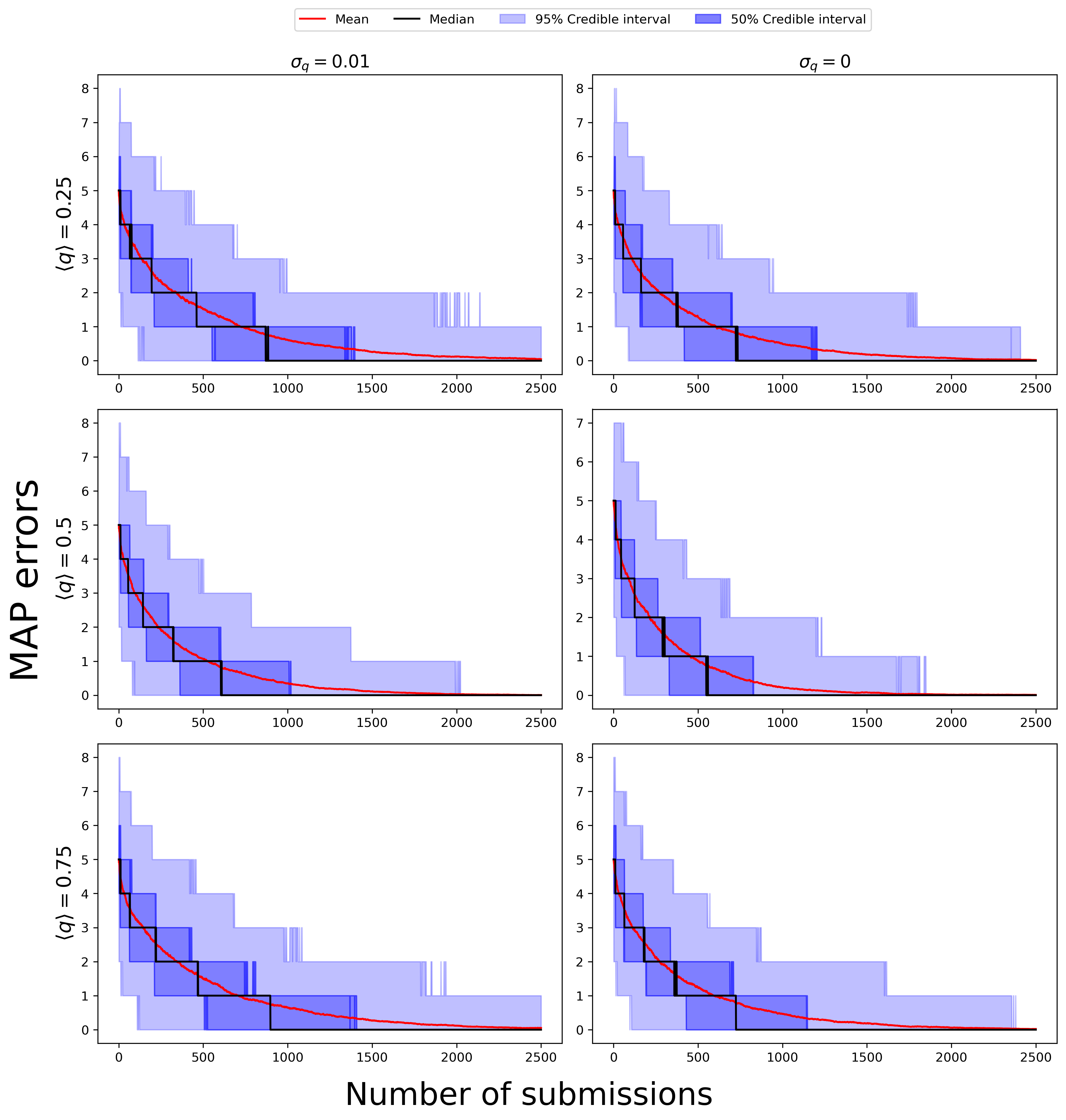}
    \caption{MAP errors comparing the distribution of variance $\sigma_q = 0.01$, as in the main text, to one of zero variance (all sampled quality factor have the same value). 
    Although the number of submissions necessary to find the correct configuration decreases when we change from $\sigma_q =0.01$ to $0$, it does not change dramatically for $\expval{q} = \nfrac{1}{2}$, in both cases needing more than $500$ submissions in the median case.
    Thus, the overall lower bound is larger than $500$ submissions and our choice of $\alpha=\beta=12$ ---  or analogously, a Beta distribution with $\expval{q} = \nfrac{1}{2}$. and $\sigma_q = 0.01$ --- as an estimator for the lower bound in a more realistic case.   }\label{fig:c_map_delta}
\end{figure}

\begin{figure}[p]
    \centering
    \includegraphics[width=.9\textwidth]{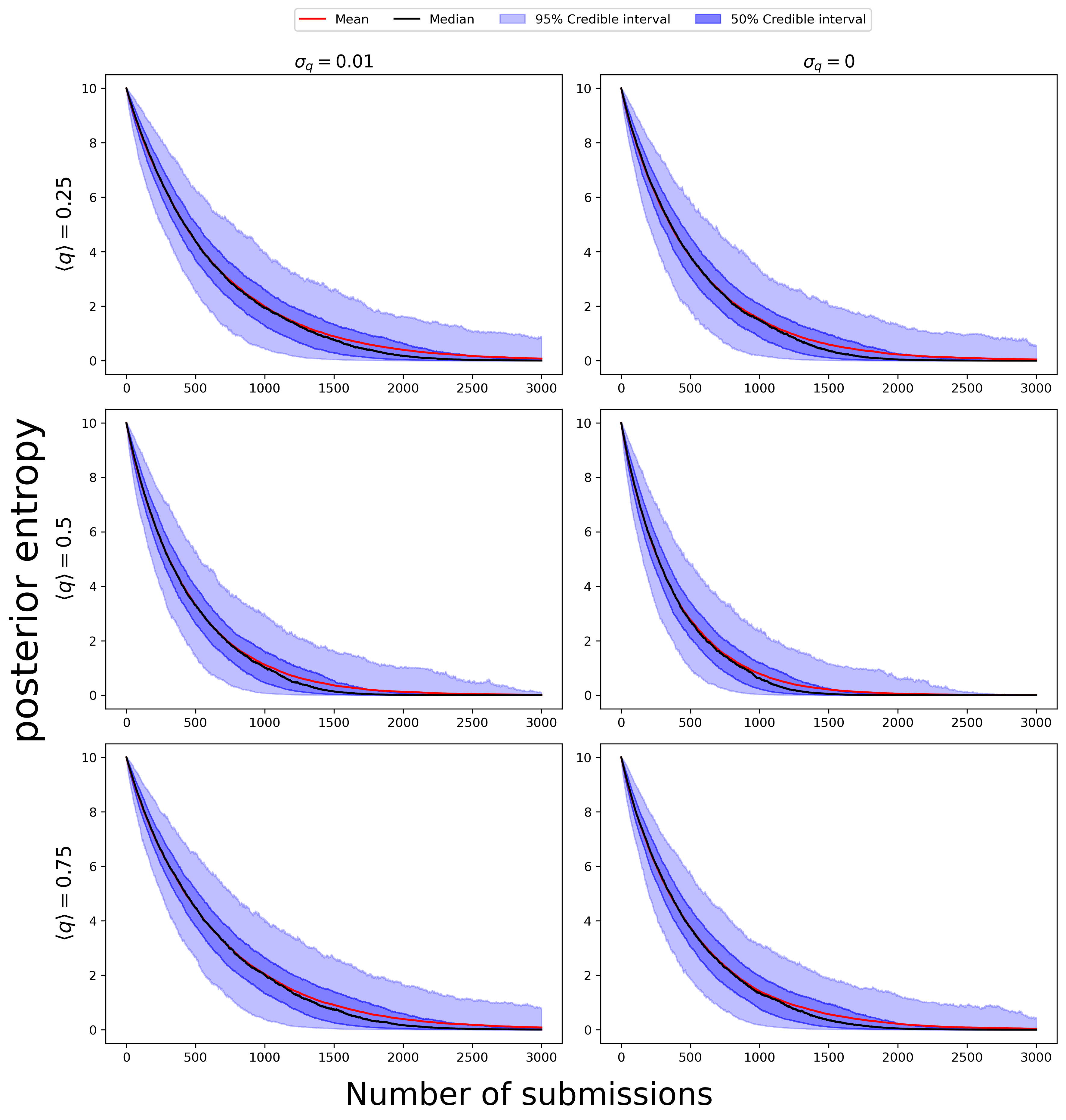}
    \caption{The posterior's entropy comparing the distribution of variance  $\sigma_q = 0.01$, as in the main text, to one of zero variance. 
    Similarly to Fig. \ref{fig:c_map_delta}, we notice that the difference is not significant for  $\expval{q} = \nfrac{1}{2}$.
    }
    \label{fig:c_entropy_delta}
\end{figure}

\begin{figure}[p]
    \centering
    \includegraphics[width=.9\textwidth]{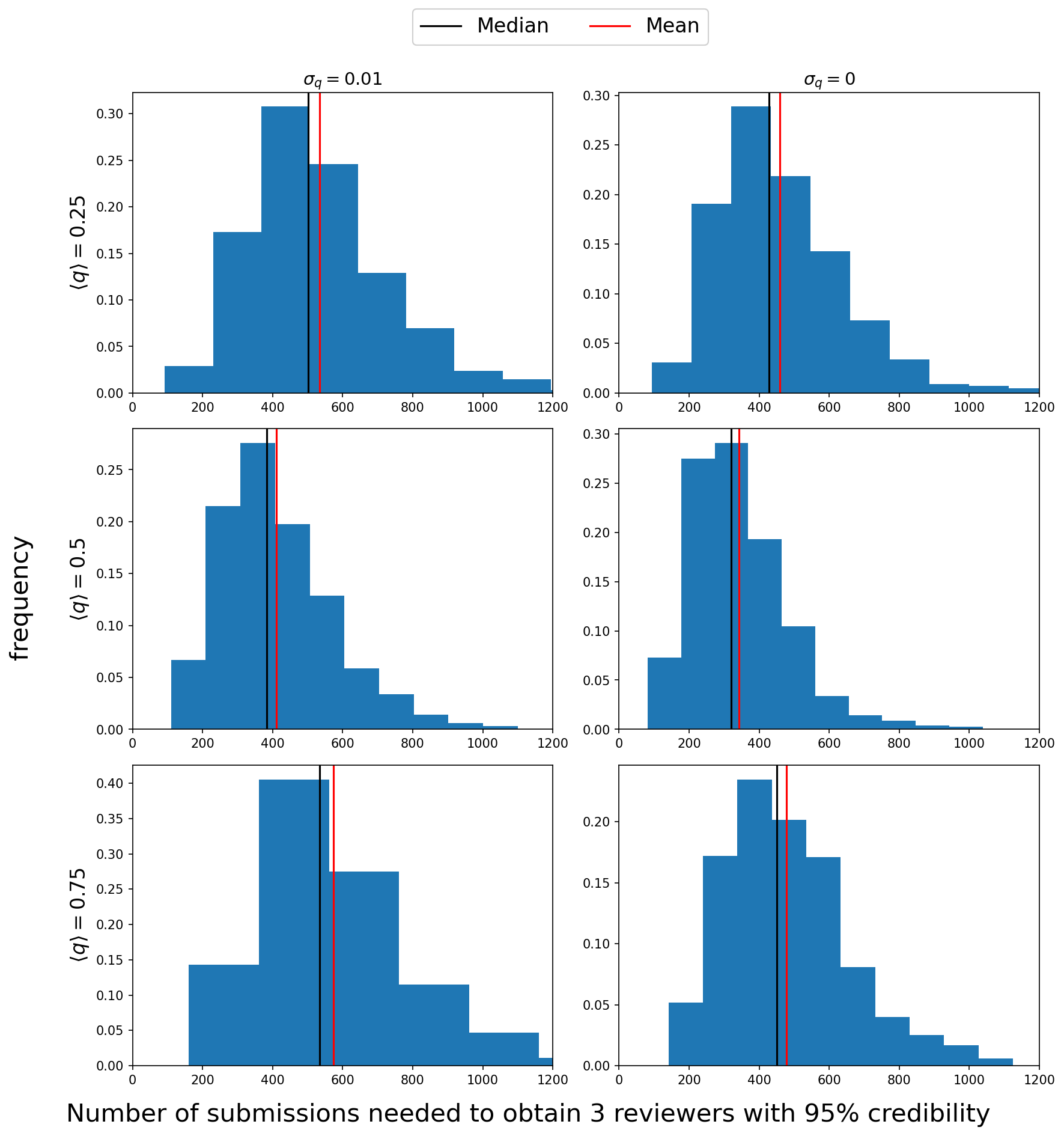}
    \caption{Number of submissions necessary to obtain $95\%$ credibility for three suggested reviewers obtained for different sampling distributions of quality factors. We notice that 300 submissions are still necessary to obtain this degree of credibility even for zero variance.
    }
    \label{fig:p3_delta}
\end{figure}

On the other hand, we also observe (in Figs. \ref{fig:C_map_different}, \ref{fig:C_entropy_different} and \ref{fig:p3_different}) that the smaller the variance, $\sigma_q$, the fewer submissions are necessary, indicating that the lower bound is also found for minimal variance in quality, $\sigma_q$. 
Although a scientist with $\sigma_q \to 0$ is not realistic, this can still be simulated in the model ---  $q_\mu = \expval{q}$ for every submission. In Fig. \ref{fig:c_map_delta} we present a comparison for MAP errors of quality $\sigma_q = 0.01$ and $\sigma_q = 0$. 
We observe that although the number of submissions necessary is smaller for $\sigma_q=0$, in the case of $\expval{q}= \nfrac{1}{2}$ the difference is not significant --- both require a little over $500$ submissions.
The analogous results for the posterior's entropy --- presented in Fig. \ref{fig:c_entropy_delta} --- shows a bigger difference but is still necessary to have more than $1500$ submissions to correctly classify reviewers.
Finally, the number of submissions necessary to reach $95\%$ with  $\sigma_q=0$  is presented in Fig. \ref{fig:p3_delta}. Even in the limit of zero variance, around 300 submissions are necessary to obtain 3 reviewers classified with $95\%$ credibility.

\newpage

\section{Errors and aggressive strategy for the fourth metric}\label{SIsec:agressive} \paragraph{}

Here, we delve deeper into the fourth metric discussed in the main text by providing a detailed analysis of the number of misclassifications occurring when attempting to achieve $95\%$ credibility among the top three reviewers, meaning how often at least one of the reviewers between the top three with higher than $95\%$ credibility are rivals in the ground truth simulation.

We also compare the number of submissions required to achieve this credibility level when randomly selecting reviewers from the set $\mathcal{R}$ with a uniform probability, as described in the main text, to a more aggressive strategy where the author utilizes information from previous submissions to suggest reviewers. This aggressive strategy involves selecting reviewers, sampling $\mathcal{S}_m$, based on the marginal posterior of previous submissions --- $\rho_i(m-1)$ as defined in \eqref{rho_def}. The first suggested reviewer is selected with a probability proportional to their marginal posterior, and this process is repeated for the remaining reviewers, excluding those that have been previously selected.

\begin{figure}[p]
    \centering
    \includegraphics[width=.8\textwidth]{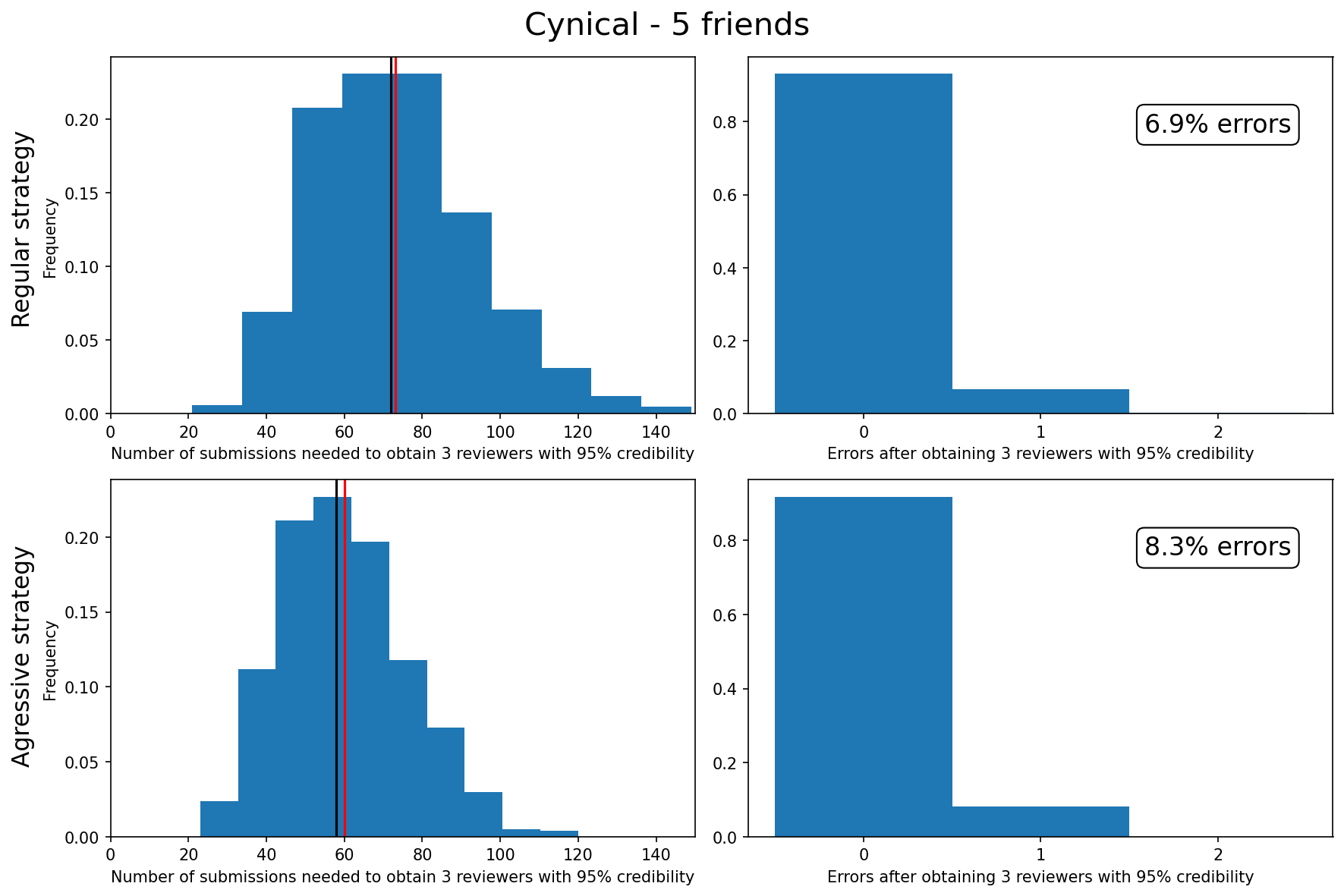}
    \vspace{-.1cm}
    \caption{Number of submissions necessary to obtain $95\%$ credibility in the three most likely reviewers to be friendly (on the left) and number of misclassified reviewers among these three (on the right). 
    These results are for the cynical model simulation with 5 (out of 10) friends in the simulation's ground truth. 
    When comparing the regular (uniform) strategy to the aggressive strategy, the number of submissions has a modest decrease. Meanwhile, the probability of having at least one misclassification increases. 
    }\label{fig:agressive_cynical_5}
    \vspace{.5cm}    
    \includegraphics[width=.8\textwidth]{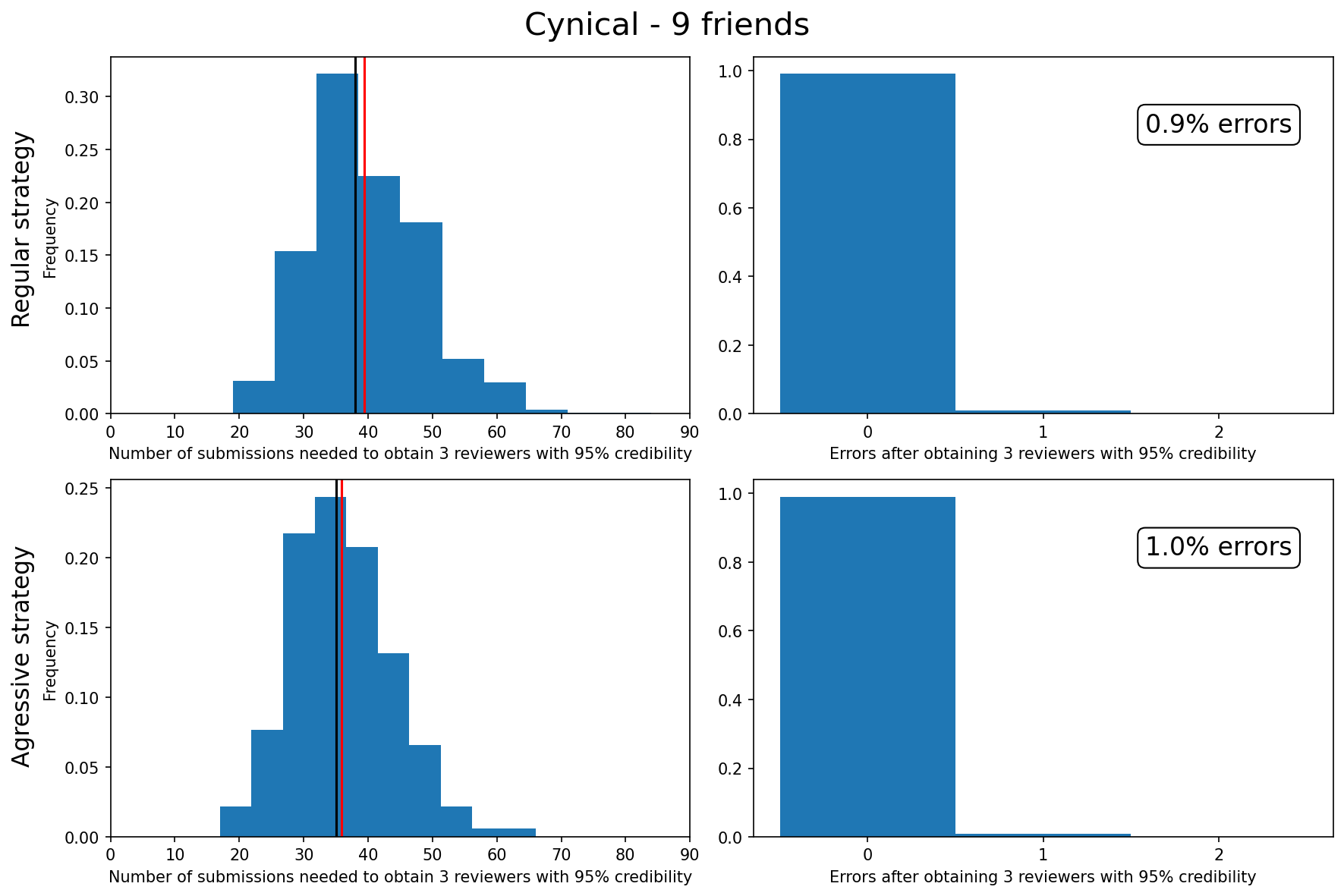}
    \vspace{-.1cm}
    \caption{Number of submissions necessary to obtain $95\%$ credibility in the three most likely reviewers to be friendly (on the left) and number of misclassified reviewers among these three (on the right). Made for the cynical model and 9 (out of 10) friends in the simulation's ground truth. When switching from the regular to the aggressive strategy, the number of submissions had a modest decrease.   
    }\label{fig:agressive_cynical_9}
\end{figure}

The number of misclassifications done when the author stops at $95\%$ credibility for the top three reviewers is presented in Fig. \ref{fig:agressive_cynical_5}. This figure is based on the cynical model simulation presented in the main text (5 out of 10 friends in the simulation's ground truth). We observe that the set of 3 highest credibility reviewers has at least one misclassified reviewer in 6.9\% of cases.
We also present analogous results when using the aggressive strategy 
to obtain $95\%$ credibility for the top three suggested reviewers in Fig.  \ref{fig:agressive_cynical_5}. It follows that although fewer submissions are necessary to obtain that credibility when using the aggressive strategy,  it comes at a trade-off of more common misclassifications (8.3\%).

\begin{figure}[p]
    \centering
    \includegraphics[width=.8\textwidth]{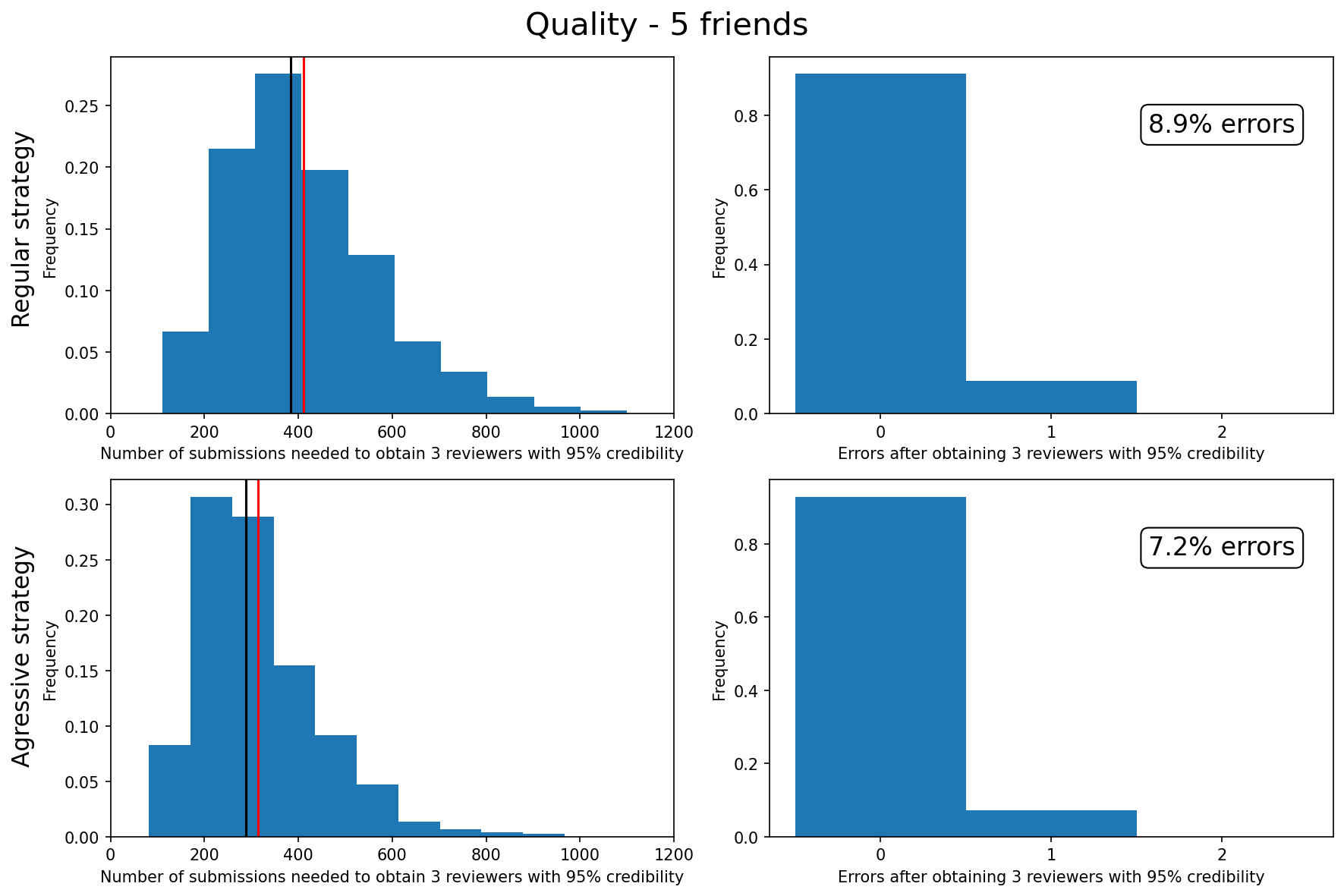}
    \vspace{-.1cm}
    \caption{Number of submissions necessary to obtain $95\%$ credibility in the three most likely reviewers to be friendly (on the left) and number of misclassified reviewers among these three (on the right). 
    These results are for the quality model simulation with 5 (out of 10) friends in the simulation's ground truth.   
    When switching from the regular to the aggressive strategy, the number of submissions had a significant decrease while the number of misclassifications descreased.
    }\label{fig:agressive_quality_5}
    \vspace{.5cm}    
    \includegraphics[width=.8\textwidth]{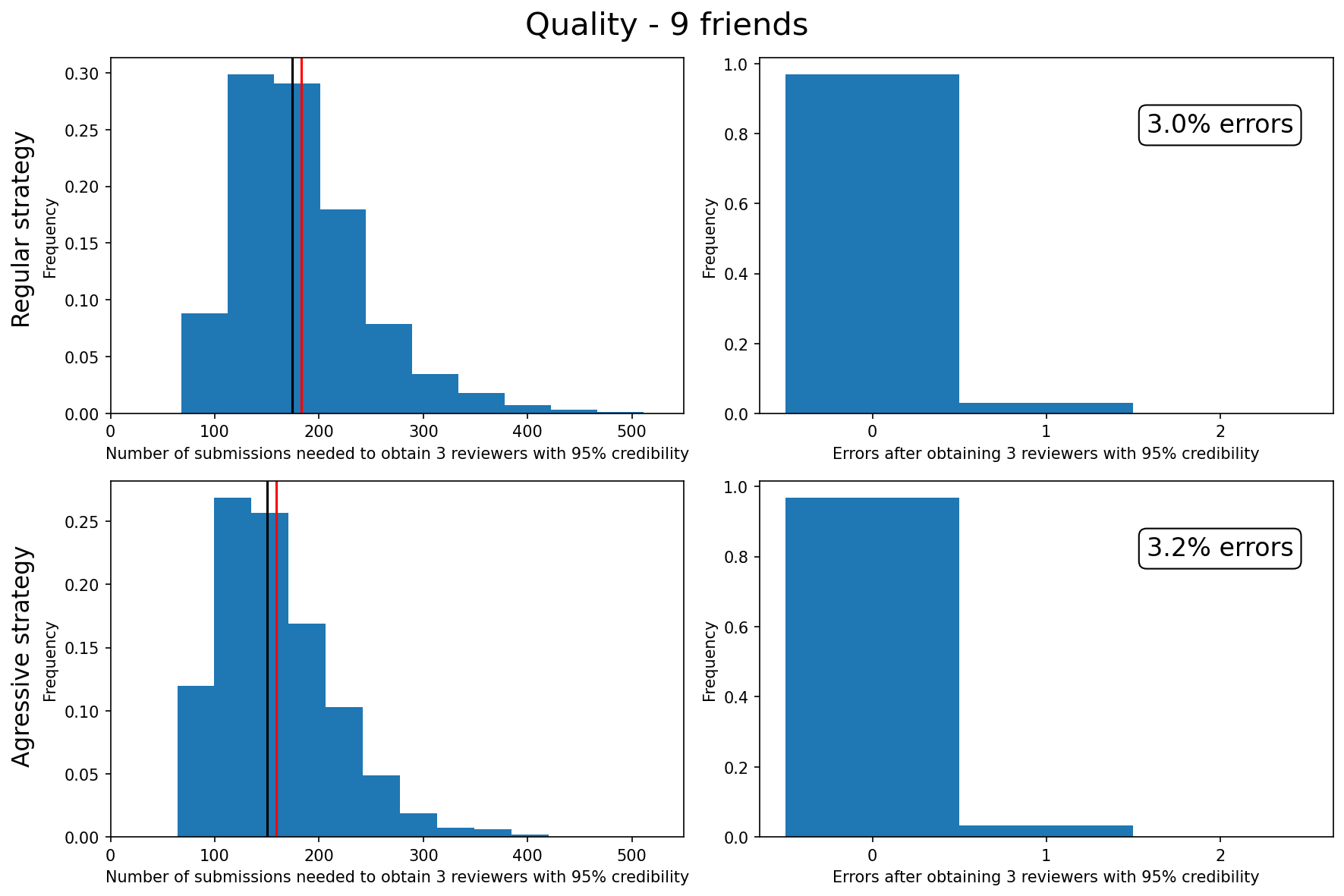}
    \vspace{-.1cm}
    \caption{Number of submissions necessary to obtain $95\%$ credibility in the three most likely reviewers to be friendly (on the left) and number of misclassified reviewers among these three (on the right). 
    These results are for the quality model simulation with 9 (out of 10) friends in the simulation's ground truth. 
    When switching from the regular to the aggressive strategy, the number of submissions had a modest decrease.
    }\label{fig:agressive_quality_9}
\end{figure}

Similar results for a ground truth with 9 friends are presented in Fig. \ref{fig:agressive_cynical_9}. When compared to the aggressive strategy, the gain in the number of submissions is modest (both around 35 submissions and similar number of mistakes). 
The results for the quality model with 5 friends are presented in Fig. \ref{fig:agressive_quality_5}. In that case, the difference in the number of submissions necessary is visible (mean value of approximately 300 in the aggressive strategy against 400 submissions) and the number of misclassification decreases (from 8.9\% to 7.2\%). The number of submissions, however, is still too large to identify a group of friendly reviewers for most researchers. In the quality model with 9 friends, presented in Fig. \ref{fig:agressive_quality_9}, a pattern similar to the cynical model emerges: the gain in the number of submissions is modest, but misclassifications increase from 3.0\% to 3.2\%.

\newpage
\newpage
